\UseRawInputEncoding
\documentclass[12pt,fleqn,twoside]{article}
\usepackage{amsmath}
\usepackage{amssymb}
\usepackage{xcolor}
\usepackage{colortbl}
\usepackage{booktabs}
\usepackage{pstricks,multido,pst-plot}
\usepackage{tikz}
\usepackage{pgfplots}
\usepackage{paralist}
\usepackage{wrapfig}
\usepackage{graphicx,epsfig}
\usepackage{marginnote}
\usepackage{todonotes}
\usepackage[margin=2cm,top=1.4cm,bottom=2.1cm,centering]{geometry}
\usepackage{tcolorbox,pgfkeys}
\definecolor{lightgray}{gray}{0.9}
\definecolor{lightyellow}{rgb}{.95,.96,.7}
\definecolor{myblue}{rgb}{.39,.54,0.82}
\definecolor{midblue}{rgb}{.7,.7,1}
\definecolor{lightblue}{rgb}{0.93,0.95,1.0}
\definecolor{mygrey}{rgb}{.90,.90,.90}
\definecolor{lightred}{rgb}{1.,0.33,0.09}
\definecolor{mygreen}{rgb}{0.40,0.75,0.16}
\definecolor{mybrown}{rgb}{0.69,0.49,0.30}
\definecolor{mypink}{rgb}{1.00,0.22,0.57}

\newtcolorbox{mybox}{colback=yellow!20!white,colframe=gray,boxrule=0.3mm,arc=3mm,outer arc=1mm}

\usepackage{enumitem}

\catcode`\@=11
\def\numberbysection{\@addtoreset{equation}{section}
     \def\theequation{\thesection.\arabic{equation}}}
\numberbysection
\def\be{\begin{equation}}
\def\ee{\end{equation}}
\newcommand\bea{\begin{eqnarray}}
\newcommand\eea{\end{eqnarray}}
\renewcommand\phi{\varphi}

\newcommand\egal{&\!\!\!=\!\!\!&}

\renewcommand{\ge}{\geqslant}
\renewcommand{\le}{\leqslant}
\def\benn{\begin{eqnarray*}}
\def\eenn{\end{eqnarray*}}

\def\Z{{\mathbb Z}}

\newcommand*\xbar[1]{%
  \hbox{%
    \vbox{%
      \hrule height 0.5pt % The actual bar
      \kern0.3ex%         % Distance between bar and symbol
      \hbox{%
        \kern-0.2em%      % Shortening on the left side
        \ensuremath{#1}%
        \kern-0.0em%      % Shortening on the right side
      }%
    }%
  }%
}

\begin{document}

%\title{\textbf{Towards a proof of the tangent method}}
\title{\textbf{Factorization in the multirefined tangent method}}

\date{}
\maketitle

\begin{center}
{\vspace{-19mm}\large \textsc{Bryan Debin}$^1$ \textsc{and Philippe Ruelle}$^1$}
\\[.8cm]
{\em {}$^1$Institut de Recherche en Math\'ematique et Physique\\ 
Universit\'e catholique de Louvain, Louvain-la-Neuve, B-1348, Belgium}
\\[.2cm]

%{\tt bryan.debin\,@\,uclouvain.be,}
%\qquad
%{\tt philippe.ruelle\,@\,uclouvain.be}
%
\end{center}

\vspace{0.5cm} 

\begin{abstract}
When applied to statistical systems showing an arctic curve phenomenon, the tangent method assumes that a modification of the most external path does not affect the arctic curve. We strengthen this statement and also make it more concrete by observing a factorization property: if $Z^{}_{n+k}$ denotes a refined partition function of a system of $n+k$ non-crossing paths, with the endpoints of the $k$ most external paths possibly displaced, then at dominant order in $n$, it factorizes as $Z^{}_{n+k} \simeq Z^{}_{n} Z_k^{\rm out}$ where $Z_k^{\rm out}$ is the contribution of the $k$ most external paths. Moreover if the shape of the arctic curve is known, we find that the asymptotic value of $Z_k^{\rm out}$ is fully computable in terms of the large deviation function $L$ introduced in \cite{DGR19} (also called Lagrangean function). We present detailed verifications of the factorization in the Aztec diamond and for alternating sign matrices by using exact lattice results. Reversing the argument, we reformulate the tangent method in a way that no longer requires an extension of the domain, and which reveals the hidden role of the $L$ function. As a by-product, the factorization property provides an efficient way to conjecture the asymptotics of multirefined partition functions.
\end{abstract}

%%%%%%%%%%%%%%%%%%%%%%%%%%%%%%%%%%%%%%%%%%%%%%%%%%%%%%%%%%%%%%%%%%%%%%%%%%%%%%%%%%%%%%%%%%%%%%%%%%%%%%%%%%
\section{Introduction}

Limit shapes have received a lot of attention over the last decade. Roughly speaking, they correspond to random geometric structures which become deterministic in an appropriate scaling limit \cite{Ok16}. Limit shapes are found in a large variety of models, including tiling models \cite{CEP96,CLP98,KO07}, large Young tableaux \cite{PR07}, sandpile models \cite{CPS08} and vertex models \cite{CP10b,CPZ10,dGKW18,DDFG20}. Special cases of limit shapes arise in models showing spatial phase transitions, by which the properties of the models, such as the entropy or the correlations, are qualitatively different in different regions. Among others, it gives rise to arctic curves, namely interfaces between frozen regions and disordered regions. The prototypical and best known example, which also gave its name to all the others, is the arctic circle for domino tilings of Aztec diamonds \cite{JPS98}. Beyond the mere determination of the arctic curves, to which the present paper is more specifically oriented, there is a number of fascinating but challenging questions, like the fluctuations around the arctic curves \cite{Jo05,JN06} or the inhomogeneous statistics inside the disordered region \cite{ADSV16,GBDJ19}.

The tangent method, developed by Colomo and Sportiello in \cite{CS16}, is a very efficient but non-rigorous method to derive analytically the shape of the arctic curve in models which possess a formulation in terms of paths. The method is based on 1-refined partition functions, for which the starting point of the most external path is moved along the boundary of the domain. One generally observes that the deformed path includes a rectilinear portion which touches the arctic curve tangentially. Taking the new starting point as a parameter, this observation yields, in principle, a one-parameter family of tangent lines whose envelope is the sought arctic curve. Although no example is known where the tangent method has been shown to fail, it relies on assumptions which can be hard to prove (see Section \ref{sec2}).

In concrete terms, the above procedure does not allow to compute, given the starting point, the line tangent to the arctic curve, because the 1-refined partition functions contain no manifest information on the shape of the deformed most external path, in particular on its slope. To circumvent the difficulty, one usually fixes a starting point outside the domain and compute the likeliest point of entry into the domain, thereby providing a second point from which the slope of the rectilinear part can be computed. 

In this paper, we make the observation that, in the scaling limit, the contribution of the outermost path to the (possibly refined) partition function factorizes from the contribution of the other paths, leading to a kind of recurrence relation $Z^{\text{1-ref}}_{n+1} \simeq Z_1^{\rm out} \, Z_n$. In addition, the one-path contribution $Z_1^{\rm out}$ of the most external path can be given an explicit expression in terms of the nature of the discrete paths present in the model and the arctic curve itself. In this expression, the Lagrangean function (or large deviation function, or entropy function) introduced in \cite{DGR19} plays a central role. By applying the factorization several times, the generalization to multirefinements is straightforward.

The factorization property has two practical and useful consequences. First, it allows to reformulate the tangent method directly in terms of 1-refined partition functions, without the need to fix a starting point outside the domain in order to compute a likeliest entry point. Second, it provides an efficient way to compute the asymptotics of multirefined partition functions from the relation $Z^{\rm multiref}_{n+m}/Z_n \simeq Z_m^{\rm out}$, explicitly computable once the arctic curve of the model is known.

The article is organized as follows. In Section \ref{sec2}, the conjectural factorization property is explained, motivated and made explicit. Many checks of the factorization property are presented in the following two sections. In Section \ref{sec3}, the factorization is verified in full generality for domino tilings of Aztec diamonds. Exact multirefined lattice partition functions are obtained, where the starting and ending points of the $m$ most external paths are moved to arbitrary points, and shown to satisfy the factorization conjecture. Moreover each of the $m$ paths contribution exactly matches the one computed with the Lagrangean function. In Section \ref{sec4}, several checks are successfully carried out, analytically or numerically, in the case of alternating sign matrices. (Many more numerical checks, not included in the present paper, have been made for the 6-vertex model, at isolated points of the disordered and antiferrroelectric phases.) The last section contains our reformulation of the tangent method on the sole basis of 1-refined partition functions. Two short appendices contain technical material.

%%%%%%%%%%%%%%%%%%%%%%%%%%%%%%%%%%%%%%%%%%%%%%%%%%%%%%%%%%%%%%%%%%%%%%%%%%%%%%%%%%%%%%%%%%%%%%%%%%%%%%%%%%
\section{The factorization property} 
\label{sec2}

The factorization we want to discuss is suggested by the analysis we have presented in \cite{DGR19} to prove the tangency property assumed in the tangent method. It has been sketched, and written explicitly in special cases, by Sportiello \cite{Sp15,Sp19} under the names of entropic tangent method and 2-refined tangent method, described as alternative and tentatively more rigorous ways to compute the shape of arctic curves. The factorization we propose appears to be more general and more explicit. 

The tangent method is believed to apply to a large variety of systems which can be formulated in terms of interacting directed lattice paths. The interaction usually takes the form of a non-intersecting or non-crossing property, but may not be restricted to those. The method distinguishes the most external path from the other paths, which yield the bulk contribution and are responsible for the formation of the arctic curve itself. It essentially relies on the two basic assumptions. First, the outermost path alone has no effect on the arctic curve, neither on its existence nor on its shape. The main argument for this is that the arctic curve is a volume effect that cannot be counterbalanced by a single path. Thus the outermost path can be modified by changing its starting or ending point, or both, without interfering with the arctic curve. This brings up the second main assumption, based on many observations: the modified external path has straight portions which connect tangentially to the arctic curve (in the scaling limit, the external path becomes a class ${\cal C}^1$ curve almost surely). It is this second assumption that makes the tangent method so efficient, allowing to determine the analytical shape of the arctic curve from a family of tangent straight lines. 

%\begin{figure}[t]
%\begin{center}
%\begin{tikzpicture}
%\clip (-0.5,-1) rectangle (25,-9);
%\draw (3.5,-5) node{\includegraphics[scale=0.35,angle=-90]{monomers_n100_m91_v6_black_monomers}};
%\draw(13.5,-5) node{\includegraphics[scale=0.35,angle=-90]{monomers_n100_m91_v6_path}};
%\end{tikzpicture}
%\end{center}
%\vspace{-1truecm}
%\caption{Typical configuration of a 1-refinement of an Aztec diamond of order $n=100$ with a displaced starting point for the outermost path. The tiling configuration is shown on the left, the corresponding path configuration on the right.}
%\label{fig1}
%\end{figure}

Though very reasonable, the first assumption is clearly the most difficult to justify in general, because its validity may depend on the interaction existing between the paths (the reference \cite{Ag20} is one of the rare cases where this was rigorously proved). But it clearly postulates a separation between the outermost path and the other paths. It served as a basis in \cite{DGR19} to address the second assumption, which was proved to hold in a fairly general setting. The analysis relied on this separation in the following way. The bulk of inner paths is responsible for the arctic curve; then the last, most external path is constrained to stay away from the arctic curve itself. It reduces the problem to analyzing, in the scaling limit, the shape of this single path conditioned to avoid the region bordered by the arctic curve, see Figure \ref{fig2} in Section \ref{sec3}. We briefly review the analysis before formulating our factorization conjecture.

When $n$, the linear size of the system, is finite, the outermost path, like all other paths, is made of a sequence of elementary steps chosen from a finite set $S$, characteristic of the model considered, starting and ending at prescribed points $P_i$ and $P_f$. Each path has its own weight, which depends on its constituent steps (and possibly on the order in which the steps are taken). As $n$ gets large, the distance between the starting and ending points is proportional to $n$, so that we set $P_i = (n a_i, n b_i)$ and $P_f = (n a_f, n b_f)$. One can then evaluate the asymptotics of the weighted sum $Z$ of all paths\footnote{The directedness of the paths ensures that the paths cannot loop around undefinitely before reaching the endpoint, and makes sure that $Z$ is finite.} between $P_i$ and $P_f$. When there are no constraints on the paths, that is, all possible paths are included, the asymptotic behaviour is exponential, 
\be
\lim_{n \to \infty} \, \frac 1n \log Z = (a_f-a_i)L(t), \qquad t = \frac{b_f - b_i}{a_f - a_i},
\label{Lfun}
\ee
and controlled by a large deviation function $L$ whose argument is the slope $t$ of the straight line joining $P_i$ to $P_f$. In case the weight of a path is the product of the weights of its elementary steps, the function $L$ is computable in terms of the steps in $S$ and their weights, see the Appendix \ref{appA} for a brief account. Under the same assumptions, it was proved that the function $L$ is strictly concave \cite{DGR19}. In the general case, the concavity of $L$ is easy to establish, but we are not aware of a general result ensuring its strict concavity, though the property has been seen to hold in a number of examples, among which the generic 6-vertex model (see Appendix \ref{appA}). The strict concavity of $L$ is crucial in what follows.

In addition to fixing the endpoints, one may consider the weighted enumeration of the paths which pass through fixed intermediate points. By increasing their number, the intermediate points form a discretization of a continuous trajectory. This leads to consider the weighted contribution $Z[f]$ of all the paths which, in the scaling limit, collapse to a fixed continuous function $f(x)$ with boundary values $f(a_i)=b_i$ and $f(a_f)=b_f$. It was found that $Z[f]$ satisfies \cite{DGR19}
\be
\lim_{n\to\infty} \frac1n \log Z[f] = S[f], \qquad S[f] = \int_{a_i}^{a_f} {\rm d}x \, L\big(f'(x)\big),
\label{Zf}
\ee
The function $L$ was also called a Lagrangean function in view of the fact that the rate $S[f]$ appears as a classical action. $Z[f]$ can be interpreted as the unnormalized probability that the path from $P_i$ to $P_f$ stays in the scaling neighbourhood of the graph of $f(x)$.

At this point, we can formulate the question raised above about the single path:  which function $f$, starting from $P_i$, ending at $P_f$ and conditioned to stay in a domain $\cal D$, maximizes $S[f]$ ? If the maximizing function $f^*$ is unique, the partition function is exponentially dominated by $Z[f^*]$,
\be
Z \simeq Z[f^*] \simeq {\rm e}^{nS[f^*]} \qquad \Longleftrightarrow \qquad \lim_{n\to\infty} \frac1n \log Z = S[f^*] = \int_{a_i}^{a_f} {\rm d}x \, L\Big(\frac{{\rm d}f^*}{{\rm d}x}\Big). %L\big({f^*}'(x)\big).
\ee 
Equivalently the relative probability that the path follows any other trajectory $f$, equal to $Z[f]/Z[f^*]$ $\simeq {\rm e}^{n(S[f]-S[f^*])}$ goes to 0 exponentially fast with $n$; in the scaling limit, the single path follows the trajectory $f^*$ almost surely. 

By using the strict concavity of $L$, it has been shown \cite{DGR19} that the maximum $f^*$ within the set of continuous and piecewise ${\cal C}^1$ functions is indeed unique, is of class ${\cal C}^1$ and corresponds to the path that minimizes the distance between the starting and the ending points while remaining in $\cal D$. As a corollary, if the domain $\cal D$ allows it, the function $f^*$ is the straight line from $P_i$ to $P_f$. If not (when the straight line would cross the boundary of $\cal D$), the function $f^*(x)$ is composed of straight segments in the interior of $\cal D$ and portions of the boundary of $\cal D$, namely pieces of the arctic curve. The tangency property assumed in the tangent method follows from the function $f^*$ being ${\cal C}^1$.

Let us come back to the full path configurations and the corresponding partition functions. We assume that the number of paths is $n$, which is directly related to the linear size of the system. As mentioned earlier, the tangent method strongly suggests a separation between the outermost path and the other paths. For convenience, we will view the most external path as a path added to a system of $n$ inner paths, which forms the core of a configuration and produces the arctic curve (we will add more external paths in a moment). The separation postulate suggests that in the scaling limit, the full partition function $Z_{n+1}$ pertaining to the configurations of $n+1$ paths factorizes into the partition function $Z_n$ relative to the $n$ core paths and the contribution $Z_1^{\rm out}$ of the outermost path conditioned not to cross the arctic curve, 
\be
\log Z^{}_{n+1} = \log Z^{}_n + \log Z^{\rm out}_1 + o(n).
\ee
We note that $\log Z^{}_{n+1}$ and $\log Z^{}_{n}$ ought to contain identical dominant terms (in particular a volume contribution proportional to $n^2$) up to order $n$, and differ by subdominant terms of order $n$, for which the contribution $Z^{\rm out}_1$ exactly compensates. So at the level of the entropies, we would expect the following identity,
\be
\lim_{n \to \infty} \frac 1n \log \frac{Z^{}_{n+1}}{Z^{}_n} = \lim_{n \to \infty} \frac 1n \log Z^{\rm out}_1.
\ee
We also note that in the full system of $n+1$ paths, the outermost one may or may not have its endpoints displaced. Such a modification, at the heart of the tangent method, will affect both $Z^{}_{n+1}$ and $Z_1^{\rm out}$, but not $Z^{}_n$. If the identity holds, the changes on $Z^{}_{n+1}$ and $Z_1^{\rm out}$ are predicted to be exactly identical, at dominant order.

The above identity is reasonable and altogether not surprising. Again the underlying idea is that the behaviour of the bulk paths is essentially unaffected by the external path, viewed as a single random path confined to a domain which is itself random at finite volume, but becomes deterministic in the scaling limit. 

$Z^{\rm out}_1$ is the scaling limit of the partition function for a single random path, constrained by the presence of the arctic curve. According to the discussion recalled above, it must be given by $Z[f^*]$, for the function $f^*$ which maximizes $S[f]$ in the given domain. We can therefore make the previous identity more explicit,
\be
\lim_{n \to \infty} \frac 1n \log \frac{Z^{}_{n+1}}{Z^{}_n} = \int_{a_i}^{a_f} {\rm d}x \, L\Big(\frac{{\rm d}f^*}{{\rm d}x}\Big),
\label{2ref}
\ee
where $Z_{n+1}$ is the partition function for a system of $n+1$ lattice paths for which the outermost path goes from $P_i$ to $P_f$.

The discussion can be readily generalized to several external paths. If adding one path does not perturb the arctic curve, adding two, three, and in general $m$ external paths does not either, provided $m$ is small. It gives rise to a $2m$-refined partition function $Z_{n+m}$, depending on $m$ starting points and $m$ ending points. The natural generalization of (\ref{2ref}) reads
\be
\lim_{n \to \infty} \frac 1n \log \frac{Z^{}_{n+m}}{Z^{}_n} = \sum_{j=1}^m S[f_j^*] = \sum_{j=1}^m \; \int_{a_{i,j}}^{a_{f,j}} {\rm d}x \, L\Big(\frac{{\rm d}f_j^*}{{\rm d}x}\Big),
\label{multi}
\ee
where $f_j^*$ is the likeliest trajectory of the $j$-th path. If the $m$-th path is the uppermost and the first one the lowermost, then the inequalities $f_1^* \le f_2^* \le \ldots \le f_m^*$ hold. The maximizing function $f^*_j$ is in principle computed with respect to its own domain ${\cal D}_j$, bordered by the graph of $f_{j-1}^*$ (that is, the artic curve if $j=1$). But in practice, the constraint from the rectlinear portions of $f_{j-1}^*$ are ineffective so that $f^*_j$ is only constrained by the arctic curve. In other words, each $f_j^*$ can be computed as in the 2-refined situation. In particular, each $S[f_j^*]$ is given by the same function $S[f^*](a_i,b_i;a_f,b_f)$ evaluated at $(a_{i,j},b_{i,j};a_{f,j},b_{f,j})$. Examples are given in Section \ref{sec3}.

The formula (\ref{multi}) is the main point of this article and clearly remains conjectural. It is expected on the basis of the arguments reviewed above, but could possibly fail, like the tangent method itself, in situations where the interactions between the lattice paths are pathological. Many checks are presented in Section \ref{sec3} and \ref{sec4}, namely for Aztec diamonds, in which the paths are strictly non-intersecting, and for alternating sign matrices, in which the paths can have kissing points. Specifically for the Aztec diamonds, the conjecture (\ref{multi}) will be verified in full generality, that is, for any $m$. In more general situations, the formula (\ref{multi}) allows to conjecture the asymptotic value of multirefined partition functions, provided the arctic curve is known.

As a special case of (\ref{2ref}), the 1-to-0 discontinuity which is the core of the 2-refined tangent method discussed in \cite{Sp15} readily follows. The 2-refined partition function is $Z_{n+1}$ computed with varying endpoints $P_i$ and $P_f$ for the external path. One considers the asymptotic value of the following ratio,
\be
\frac{Z_{n+1}}{Z_n \, Z[f_{\rm st}]} \simeq \exp{\big\{n \big(S[f^*]-S[f_{\rm st}]\big)\big\}},
\ee
where $f_{\rm st}$ is the straight trajectory between $P_i$ and $P_f$. If $P_i$ and $P_f$ are varied so that the straight line $f_{\rm st}$ does not cross the arctic curve, then the maximizing $f^*$ {\it is} the straight line $f_{\rm st}$ and the ratio is of order 1. The limiting values of $P_i$ and $P_f$ for this to be true are such that the line $f_{\rm st}$ is tangent to the arctic curve. For $P_i$ and $P_f$ beyond these limiting values, the maximizing $f^*$ is no longer the straight line, with the consequence that $S[f^*] < S[f_{\rm st}]$. It follows that the ratio becomes exponentially small in $n$ and goes to 0 in the scaling limit. The previous formula provides an explicit\footnote{That expression may not look so explicit. We will see in the following sections that it can be made completely explicit in concrete examples, provided the shape of the arctic curve is known.} expression for the exponential rate, namely $S[f^*]-S[f_{\rm st}]$.

The observation underlying the entropic tangent method \cite{Sp19}, which is based on a 1-refined partition function, also follows from the formula (\ref{2ref}) and the nature of the maximizing trajectory $f^*$.

%%%%%%%%%%%%%%%%%%%%%%%%%%%%%%%%%%%%%%%%%%%%%%%%%%%%%%%%%%%%%%%%%%%%%%%%%%%%%%%%%%%%%%%%%%%%%%%%%%%%%%%%%%
\section{Multirefinements for Aztec diamonds}
\label{sec3}

Domino tilings of Aztec diamonds are among the first and most studied tiling models \cite{EKLP92,CEP96,JPS98}. Figure 1 shows on the left one of the $2\,097\,152$ possible tilings of the Aztec diamond of order 6. For a general order $n$, the number of tilings is equal to $2^{n(n+1)/2}$.

\begin{figure}[h]
\begin{center}
\includegraphics[scale=1]{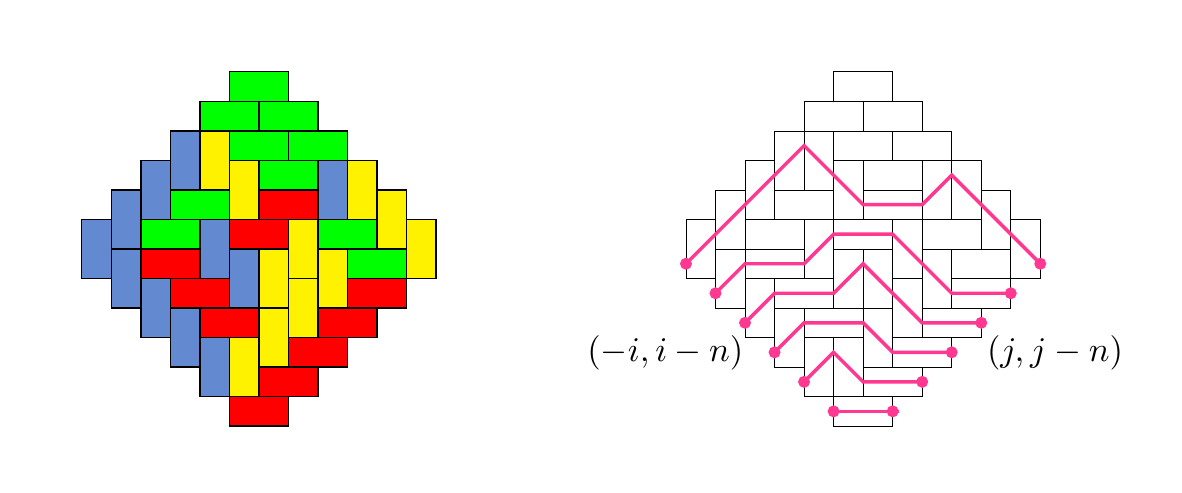}
\end{center}
\vspace{-01cm}
\caption{The bijection between a tiling of the Aztec diamond of order 6 and the corresponding 6-uple of non-intersecting paths. The two kinds of horizontal and vertical dominos are distinguished by the parity of their coordinates.}
\label{fig1}
\end{figure}

We decide to weigh the dominos according to their orientation, by assigning the horizontal and vertical dominos a weight $\sqrt{w}$ and 1 respectively. The weight of a tiling is then equal to $w^{N_h/2}$ with $N_h$ the number of horizontal dominos. The partition function for the Aztec diamond of order $n$ is given by
\be
Z_n = \sum_{\rm tilings} w^{N_h/2},
\ee
In this section, the notation $Z_n$ will be used exclusively for the standard tilings, equivalently for the configurations of $n$ paths with the fixed endpoints as shown in Figure \ref{fig1}. 

The domino tilings of the Aztec diamond of order $n$ are in bijection with the $n$-uples of non-intersecting Schr\"oder-like paths \cite{LRS01,EF05}, see Figure \ref{fig1}, where the endpoints of the paths are fixed on the lower left and lower right boundaries, as shown\footnote{This is one possible choice. There are in general several ways to describe the configurations of a given model in terms of paths.}. We choose the coordinate system for which the center of the diamond is at $(\frac 12,\frac 12)$ and such that the starting and ending points of the paths have coordinates $p_i \equiv (-i,i-n)$ and $q_j \equiv (j,j-n)$, for $i,j=1,\ldots,n$. The paths themselves are made of three elementary steps: $(1,1)$, $(1,-1)$ and $(2,0)$ carried respectively by the blue, yellow and red dominos (the lowest path from $p_1$ to $q_1$ cannot contain a down step). To get the correct weight, we give the diagonal steps $(1,1)$ and $(1,-1)$ a weight 1, and the level step $(2,0)$ a weight $w$ (there is an equal number of red dominos, carrying a level step, and green dominos, which carry no step at all). 

The Lindstr\"om-Gessel-Viennot lemma \cite{Li73,GV85} allows to compute the partition function in terms of the path configurations. Let $A_{i,j}$ be the weighted sum over the unconstrained paths from $p_i$ to $q_j$ and made of the above three steps. Then the partition function is
\be
Z_n = \det (A_{i,j})_{\,0 \le i,j \le n},
\ee
because the determinant exactly retains those path configurations with no intersections. To account for the restriction that the lowest path cannot go down, the values $i=0$ and $j=0$ have been included ($A_{i,0} = A_{0,j} = 1$ for $i,j \ge 0$).

The matrix elements $A_{i,j}$ are given in terms of trinomial coefficients,
\be
A_{i,j} = \sum_{p=0}^{\min(i,j)} \; w^p \: {i+j-p \choose i-p,j-p,p}, \qquad i,j \ge 0,
\ee
with $i-p$, $j-p$ and $p$ the numbers of down, up and horizontal steps respectively. To compute refined partition functions, the use of LU decompositions is particularly convenient \cite{DFL18}. Using generating functions, the LU decomposition of the semi-infinite matrix $(A_{i,j})_{i,j \ge 0}$ was obtained in \cite{DFL18} for $w=1$, but the generalization to any $w$ is straightforward. The result is
\be
L_{i,j} = {i \choose j} \quad (i \ge j), \qquad
U_{i,j} = (1+w)^i \, {j \choose i} \quad (i \le j).
\label{LU}
\ee
The semi-infinite LU decomposition directly yields that of any principal submatrix, by restriction; it also turns out to be crucial for the computation of general refined partition functions.

The calculation of $Z_n$ is now easy,
\be
Z_n = \det (U_{i,j})_{\,0 \le i,j \le n} = \prod_{i=0}^n \: U_{i,i} = (1+w)^{n(n+1)/2}.
\label{Zn}
\ee
Before proceeding to the verification of the factorization formula (\ref{multi}), we briefly discuss the scaling limit.

\begin{figure}[t]
\begin{center}
\includegraphics[scale=1]{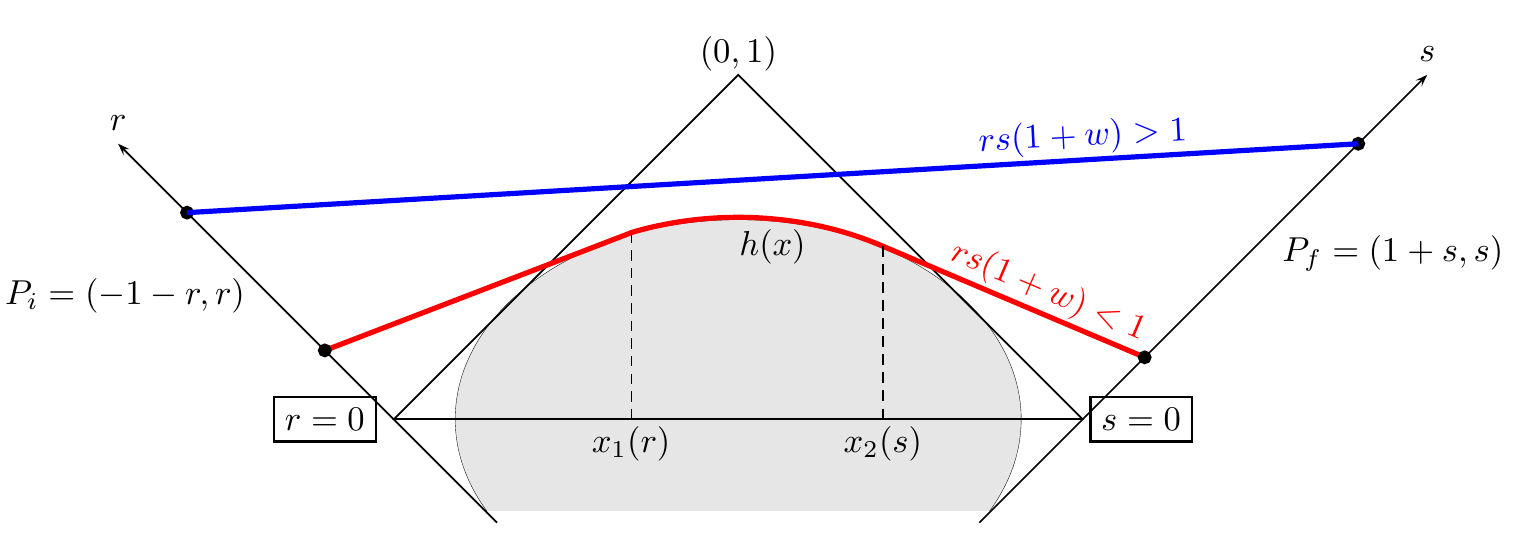}
\end{center}
\vspace{-0.5cm}
\caption{Graphical representation of a 2-refinement for the Aztec diamond. An extra path is added, starting from the left boundary and ending on the right boundary. In the scaling limit, the random path collapses almost surely on a trajectory that looks like the blue straight line or the red curve, depending on the initial and final positions.}
\label{fig2}
\end{figure}

The scaling limit is obtained by dividing all distances by $n$ and taking the limit $n$ going to infinity. The Aztec diamond goes to a square centered at the origin and rotated by 45 degrees, partly shown in Figure \ref{fig2}. For generic $w$, the arctic curve is an ellipse, with equation \cite{CEP96}
\be
\frac{x^2}w + y^2 = \frac 1{1+w}.
\ee
The interior of the ellipse, namely the temperate region, is represented in Figure \ref{fig2} as the shaded area.
 
The $2m$-refinements we consider consist in imposing constraints on the $m$ uppermost paths. For convenience, we choose to keep $n$ unconstrained paths, so that the normalizing partition is always the same, and equal to $Z_n$. Different refinements can be considered, depending on the starting and ending points of the extra paths. 
The ensuing calculations in the scaling limit are not very sensitive to the actual choice, but the lattice calculations are. A natural choice is to constrain the $m$ paths to take their first step not equal to $(1,1)$ at prescribed positions $a_j$ along the upper left boundary, and likewise to constrain them to have their last step not equal to $(1,-1)$ at some other positions $b_j$ along the upper right boundary. Equivalently, we can actually move the starting and ending points from their original positions to new locations $a_j$ and $b_j$ by inserting monomers at appropriate positions. 

A simplifying choice turns out to let the $m$ paths start and end on the lines which form the continuations of the lower left and right boundaries of the diamond. More precisely, for two growing sequences $0 \le r_1 \le \ldots \le r_m$ and $0 \le s_1 \le \ldots \le s_m$, the $j$th path starts from the point $(-1-r_j,r_j)$ and ends at $(1+s_j,s_j)$, after the rescaling of the coordinates by $n$, as shown in Figure \ref{fig2}. Thus the extra path with label 1 is the innermost, that with label $m$ the outermost. The refined partition function, corresponding to the $n+m$ paths, will be denoted by $Z_{n+m}(r_1,r_2,\ldots|s_1,s_2,\ldots)$.

This second way of defining the refinements is not really different from the first one. In the scaling limit, the paths starting at $(-1-r_j,r_j)$ and ending at $(1+s_j,s_j)$ will enter and exit the domain at certain points $a_j$ and $b_j$ respectively, making contact with the first refinements. Fixing one set of points or the other is merely a change of variables. The only difference is the contributions coming from the straight portions located outside the domain, which can easily be taken into account.

Let us first consider the case of a single extra path, with starting and ending points $P_i$ and $P_f$ specified by a pair $(r,s)$. For $r$ and $s$ large enough, the straight line from $P_i$ to $P_f$ does not cross the arctic ellipse. From the discussion of the previous section, it follows that in the scaling limit, the shape of the extra path is almost surely the straight line. This case is represented by the blue line in Figure \ref{fig2}. When $r$ and/or $s$ decrease, the straight line will at some point cross the arctic ellipse. Then the path follows almost surely the trajectory represented in red in Figure \ref{fig2}, which avoids the temperate region by minimizing the length from $P_i$ to $P_f$. The factorization conjectured in the previous section implies the following limit,
\be
S(r,s) \equiv \lim_{n \to \infty} \frac 1n \log \frac{Z_{n+1}(nr|ns)}{Z_n} = S[f^*] = \int_{-1-r}^{1+s} {\rm d}x \, L\Big(\frac{{\rm d}f^*}{{\rm d}x}\Big),
\label{entropy}
\ee
for the appropriate function $L$, given below, and where $f^*$ is the blue or red trajectory, computable in terms of $r,s$. For the red trajectory, the knowledge of the arctic curve is required.

In case we add $m$ paths, and depending on the values of the pairs $(r_j,s_j)$, some of the lower paths will follow a portion of the arctic curve in the scaling limit, like the red path in Figure \ref{fig2}, while the others will be straight lines. Each of them has an associated action given by $S[f_j^*] = S(r_j,s_j)$, and the following limit should hold,
\be
\lim_{n \to \infty} \frac 1n \log \frac{Z_{n+m}(nr_1,\ldots,nr_m|ns_1,\ldots,ns_m)}{Z_n} = S(r_1,s_1) + S(r_2,s_2) + \ldots + S(r_m,s_m).
\label{mref}
\ee
%Let us note that all rectilinear parts of the trajectories $f_j^*$ must have a slope in $[-1,1]$, since otherwise they cannot be realized in terms of the three elementary steps of the underlying discrete paths.

%%%%%%%%%%%%%%%%%%%%%%%%%%%%%%%%%%%%%%%%%
\subsection{The 2-refined entropy function}

The function $S(r,s)$ appears as the entropy associated to a single path, or as the action $S[f^*]$ computed along the maximizing trajectory. To compute it, it is sufficient to have an explicit formula for $f^*(x)$ and to know the function $L$. 

As shown by (\ref{Lfun}), the function $L$ only depends on the nature of the lattice paths, through the elementary steps that make up the paths, and their weights. Its computation is reviewed in Appendix A. For the Aztec diamonds the set of elementary steps is $S=\{(1,1),(1,-1,),(2,0)\}$ with weights $1,1,w$. The corresponding function $L$ is equal to 
\be
L(t) = \log\frac{\sqrt{1+wt^2} + \sqrt{1+w}}{\sqrt{1-t^2}} + t \log\frac{\sqrt{1+wt^2} - t\sqrt{1+w}}{\sqrt{1-t^2}}, 
\qquad -1 \le t \le 1.
\ee

We first consider the case when $r$ and $s$ are small enough so that the trajectory $f^*$ touches the arctic curve. It is then composed of two line segments with slopes $t_1(r)$ and $t_2(s)$ and of a portion of arctic curve between the two tangency points $\big(x_1(r), h(x_1(r))\big)$ and $\big(x_2(s), h(x_2(s))\big)$, with $h(x)$ the arctic curve, see Figure \ref{fig2}. It is not hard to show that
\be
t_1(r)=\frac{1-r^2(1+w)}{1+r(2+r)(1+w)}, \qquad x_1(r)=-{w t_1(r)}{\sqrt{(1+w)(1+wt_1(r))}},
\ee
while $t_2(s) = -t_1(s)$ and $x_2(s) = -x_1(s)$ are obtained by symmetry. When $r$ and/or $s$ get larger, the portion of arctic curve spanned by $f^*$ decreases, until it is reduced to a single point when $t_1=t_2$, implying $rs=\frac{1}{1+w}$. This condition is therefore the borderline between the two cases illustrated in Figure \ref{fig2}. 

When $rs \le \frac1{1+w}$, the action takes a surprisingly simple form,
\bea
S[f^*] \egal (x_1+1+r)L(t_1) + \int_{x_1}^{x_2} {\rm d}x \; L\big(h'(x)\big) + (1+s-x_2) L(t_2)\nonumber \\
\egal (1+r) \log(1+r)-r \log r + (1+s) \log(1+s) - s \log s + \log(1+w).
\eea
%where we used the identity 
%\small
%\bea
%\int_{x_1}^{x_2} {\rm d}x \, \left\{ \log\left[\frac{1+\sqrt{(w+1)(1-\frac{1+w}w x^2)}}{\sqrt{1-(\frac{1+w}w)^2 x^2}}\right] - \frac{x\sqrt{1+w}}{2w\sqrt{1-\frac{1+w}w x^2}} \log\left[\frac{1+\frac{1+w}w x}{1-\frac{1+w}w x}\right] \right\} \nonumber\\
%\noalign{\medskip}
%&& \hspace{-14.3cm} = \frac{w}{1+w} \int_{\frac{w+1}{w}x_1}^{ \frac{w+1}{w} x_2} {\rm d}s \, \left\{ \log\left[\frac{1+\sqrt{(w+1)(1-\frac w{1+w} s^2)}}{\sqrt{1-s^2}}\right] - \frac s{2\sqrt{1+w-ws^2}} \log\left[\frac{1+s}{1-s}\right] \right\} \nonumber\\
%\noalign{\medskip}
%&& \hspace{-14.3cm}  = \left[ \frac 12 \Big(-1 + \frac{\sqrt{1+w-ws^2}}{1+w}\Big) \log\frac{1+s}{1-s} + \frac{sw}{1+w} \log\frac{1+\sqrt{1+w-ws^2}}{\sqrt{1-s^2}}   + \frac12 \log\frac{1+w+sw+\sqrt{1+w-ws^2}}{1+w-sw+\sqrt{1+w-ws^2}}\right]_{\frac{w+1}{w}x_1}^{ \frac{w+1}{w} x_2}. \nonumber\\
%\noalign{\medskip}
%\label{eq_integral_L2}
%\eea
%\normalsize

In case $rs \ge \frac1{1+w}$, the trajectory is a straight line of slope $\frac{s-r}{2+r+s}$, so that the integral giving $S[f^*]$ is straightforward. 

We thus obtain 
\be
S(r,s) = \begin{cases}
\displaystyle   (r+s+2) \, L\Big(\frac{s-r}{r+s+2} \Big), & rs \ge \frac1{1+w},\phantom{xx}\\ 
\noalign{\medskip}
\displaystyle  (r+1) \log(r+1) - r \log r + (s+1) \log(s+1) - s \log s + \log(1+w), & rs \le \frac1{1+w}.
\end{cases} 
\label{Srs}
\ee

Let us already note that the factorization is satisfied in the case $r=s=0$. Indeed $Z_{n+1}(0|0) = Z_{n+1}$ is the large $n$ limit of the partition function of the Aztec diamond of order $n+1$, and the conjecture (\ref{mref}) reads
\be
\lim_{n \to \infty} \frac 1n \log \frac{Z^{}_{n+1}}{Z^{}_n} = S(0,0) = \log(1+w).
\ee
The identity is satisfied in view of $Z_n = (1+w)^{n(n+1)/2}$. The same readily holds for the $2m$-refined partition function $Z_{n+m}(0,\ldots,0|0,\ldots,0)$. 

We now proceed to verify the factorization conjecture for a general $2m$-refinement, starting with the case $m=1$.

%%%%%%%%%%%%%%%%%%%%%%%%%%%%%%%%%%%%%%%%%
\subsection{Lattice 2-refined partition functions}

In terms of paths, the lattice 2-refined partition function is defined from the paths configurations of the Aztec diamond of order $n$ to which we add one extra path, with starting and ending points $p_{n+k} = (-n-k,k)$ and $q_{n+\ell} = (n+\ell,\ell)$ for $k,\ell \ge 1$. The corresponding partition function, which we denote by $Z_{n+1}(k|\ell)$, is the weighted sum of all configurations of $n+1$ non-intersecting paths starting from $\{p_1,p_2,\ldots,p_n,p_{n+k}\}$ and ending at $\{q_1,q_2,\ldots,q_n,q_{n+\ell}\}$. In the scaling limit, when $k,\ell$ are proportional to $n$, the lattice and continuous variables are related by $k=rn$ and $\ell=sn$.

\begin{figure}[t]
\begin{center}
\includegraphics[scale=1]{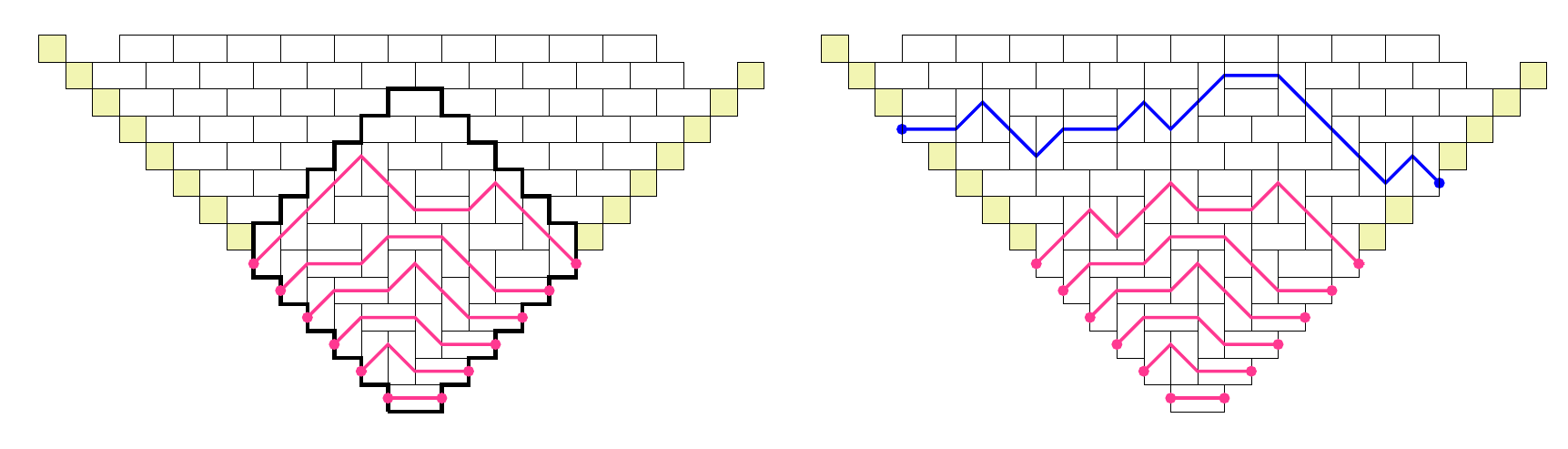}
\end{center}
\vspace{-1cm}
\caption{Graphical representation of how to add an extra path, with fixed starting and ending points, to the standard set of $n$ paths for an Aztec diamond of order $n$ ($n=6$ here). The yellow cells represent the monomers.}
\label{fig3}
\end{figure}

In terms of tilings, the way an extra path can be added is depicted in Figure \ref{fig3}. Before adding the extra path, the Aztec diamond of order $n$ is embedded in a larger quadrant (the size of which depends on the location of the path to be added) by placing one layer of monomers along the left boundary and one along the right boundary, and filling the region above the diamond with green dominos. This makes sure that the extended domain contains no more paths than the original Aztec diamond. Once the extension is realized, one removes one monomer on the left and one on the right, at the required positions; a new path, in blue in Figure \ref{fig3}, is then automatically created. 

According to the LGV formula, the 2-refined partition function is given by the following determinant of rank $n+2$,
\be
Z_{n+1}(k|\ell) = \det\big[A_{i,j}(k|\ell)\big]_{0 \le i,j \le n+1}, \quad 
A_{i,j}(k|\ell) = \begin{cases}
A_{i,j} & {\rm for\ } i,j \le n,\\
A_{n+k,j} & {\rm for\ } i=n+1,\,j \le n,\\
A_{i,n+\ell} & {\rm for\ } i \le n,\,j=n+1,\\
A_{n+k,n+\ell} & {\rm for\ } i,j=n+1.
\end{cases}
\ee

Defining the following combination of the $L$ and $U$ matrix entries given in (\ref{LU}),
\be
g = \sum_{j=1}^{\min(k,\ell)} \: L_{n+k,n+j} \, U_{n+j,n+\ell},
\ee
one can verify, by using the semi-infinite LU decomposition of $A$, that the LU decomposition of $A(k|\ell)=L(k|\ell)U(k|\ell)$ is given by
\be
L_{i,j}(k|\ell) = \begin{cases}
L_{i,j} & {\rm for\ } i\le n, j\le n+1,\\
L_{n+k,j} & {\rm for\ } i=n+1, j \le n,\\
1 & {\rm for\ } i=j=n+1,
\end{cases}
\quad 
U_{i,j}(k|\ell) = \begin{cases}
U_{i,j} & {\rm for\ } i\le n+1, j\le n,\\
U_{i,n+\ell} & {\rm for\ } i \le n, j=n+1,\\
g & {\rm for\ } i=j=n+1.
\end{cases}
\ee 
From this, we readily obtain
\be  
Z_{n+1}(k|\ell) = g \, \det\!\big[U_{i,j}(k|\ell)\big]_{0 \le i,j \le n} = g \, \det\!\big[U_{i,j}\big]_{0 \le i,j \le n},
\ee
and therefore
\be
Z_1^{\rm out}(k|\ell) = \frac{Z_{n+1}(k|\ell)}{Z_n} = g = \sum_{j=1}^{\min(k,\ell)} \: (1+w)^{n+j} \, {n+k \choose n+j} \, {n+\ell \choose n+j}.
\ee

In order to make contact with the function $S(r,s)$, we set $k=rn$, $\ell=sn$, and perform a saddle point analysis of the previous sum, for which we may assume $k \le \ell$ by symmetry in $k,\ell$. Keeping the exponential terms only, we find
\bea
&& \hspace{-1cm} g \simeq \int_0^r {\rm d}\xi \exp\Big\{n \big[(\xi+1) \log(1+w) - 2(\xi+1)\log(\xi+1) + (r+1) \log(r+1)  \nonumber\\
&& \hspace{3cm} +\, (s+1)\log(s+1) - (r-\xi)\log(r-\xi) - (s-\xi)\log(s-\xi) \big]\Big\} \nonumber\\
&& \hspace{-7mm} \simeq \exp\big\{n F(\xi^*;r,s)\big\},
\eea
where $F(\xi;r,s)$ is the function within the square brackets, and $\xi^*$ is the location of its maximum in the integration domain (if unique). The function $F$ is strictly concave over the entire domain $[0,r]$, and its derivatives at the endpoints are equal to $\partial_\xi F(0;r,s) = \log\big[rs(1+w)\big]$ and $\partial_\xi F(r;r,s) = -\infty$. The extremum condition $\partial_\xi F=0$ yields the quadratic equation
\be
(\xi^*+1)^2 = (1+w)(r-\xi^*)(s-\xi^*).
\label{quad}
\ee

Let us first assume $rs(1+w) \le 1$. Since $\partial_\xi F(0;r,s) \le 0$, the concavity of $F$ implies that it is strictly decreasing on $[0,r]$. Its maximum on the domain is therefore at $\xi^*=0$, with the value
\be
F(\xi^*;r,s) = (r+1) \log(r+1) - r \log r + (s+1) \log(s+1) - s \log s + \log(1+w),
\ee
exactly equal to $S(r,s)$.

When $rs(1+w) > 1$, the derivative of $F$ at $\xi=0$ being strictly positive, there is a unique maximum in the interior of $[0,r]$. Because the product of the two roots of (\ref{quad}) is equal to $\big(rs(1+w)-1\big)/w$, hence positive, one is in $[0,r]$ and the other is strictly larger than $r$. Thus the location of the maximum of $F$ in $[0,r]$ is at the smallest of the two roots,
\be
\xi^* = \frac 1{2w} \Big[2+(r+s)(1+w) - \sqrt{\big[2+(r+s)(1+w)\big]^2 + 4w \big[1 - rs(1+w)\big]}\Big].
\ee
Using the quadratic equation satisfied by $\xi^*$, the maximal value of $F$ can be written as
\bea
F(\xi^*;r,s) \egal (r+1) \log\frac{r+1}{r-\xi^*} + (s+1) \log\frac{s+1}{s-\xi^*} \nonumber\\
\egal - \, (r+s+2) \log\Big[\frac{(r-\xi^*)(s-\xi^*)}{(r+1)(s+1)}\Big]^{1/2} - (s-r) \log\Big[\frac{(r+1)(s-\xi^*)}{(r-\xi^*)(s+1)}\Big]^{1/2}
\label{F}
\eea
This last expression can be identified with the function $S(r,s)$, namely
\be
(r+s+2) \, L(t) = -(r+s+2) \log x(t) - (s-r) \log y(t), \qquad t=\frac{s-r}{r+s+2},
\label{L}
\ee
provided the arguments of the logarithms match. The two functions $x(t)$ and $y(t)$, discussed in the Appendix \ref{appA}, are the only positive solutions of two algebraic equations, which read in this case,
\be
xy + \frac xy + wx^2 = 1, \qquad xy - \frac xy = t(1 + wx^2).
\ee
The equality of (\ref{F}) and (\ref{L}) thus follows if one can show that the following two functions,
\be
X = \Big[\frac{(r-\xi^*)(s-\xi^*)}{(r+1)(s+1)}\Big]^{1/2}, \qquad Y = \Big[\frac{(r+1)(s-\xi^*)}{(r-\xi^*)(s+1)}\Big]^{1/2},
\ee
satisfy the same algebraic equations, which is completely straightforward by using the quadratic equation satisfied by $\xi^*$.

We have shown that 
\be
\lim_{n \to \infty} \frac 1n \log \frac{Z_{n+1}(nr|ns)}{Z_n} = F(\xi^*;r,s) = S(r,s),
\ee
and confirmed the conjecture for the 2-refined case.

%%%%%%%%%%%%%%%%%%%%%%%%%%%%%%%%%%%%%%%%%
\subsection{Lattice multirefined partition functions}
\label{multiref}

For the general multirefined case, we include $m$ additional paths to the $n$ paths forming the standard Aztec diamonds of order $n$, with $m = {\cal O}(1)$. Following the conventions used before, we set the starting points to $\{p_1,p_2,\ldots,p_n,p_{n+k_1},\ldots,p_{n+k_m}\}$ and the ending points to $\{q_1,q_2,\ldots,q_n,q_{n+\ell_1},\ldots,q_{n+\ell_m}\}$ for two strictly increasing sequences $1 \le k_1 < \ldots < k_m$ and $1 \le \ell_1 < \ldots < \ell_m$. The corresponding partition function is denoted by $Z_{n+m}(k_1,\ldots,k_m|\ell_1,\ldots,\ell_m)$ or the shorter $Z_{n+m}(\vec k|\vec \ell)$.

It is given by the determinant of the $(n+1+m)\times(n+1+m)$ matrix
\be
A_{i,j}(\vec k|\vec \ell) = \begin{cases}
A_{i,j} & {\rm for\ } i,j \le n,\\
A_{n+k_a,j} & {\rm for\ } i=n+a,\,j \le n,\\
A_{i,n+\ell_b} & {\rm for\ } i \le n,\,j=n+b,\\
A_{n+k_a,n+\ell_b} & {\rm for\ } i=n+a,\, j=n+b,
\end{cases}
\qquad 1 \le a,b \le m.
\ee

Let us write the LU decomposition of $A(\vec k|\vec \ell)$ in the following form,
\be
L(\vec k|\vec \ell) = \left[
\begin{matrix}
L & 0 \\
M & l
\end{matrix}
\right], \qquad 
U(\vec k|\vec \ell) = \left[
\begin{matrix}
U & V \\
0 & u
\end{matrix}
\right], \qquad
A(\vec k|\vec \ell) = \left[
\begin{matrix}
LU & LV \\
MU & MV + lu
\end{matrix}
\right],
\ee
where $L,U$ decompose the principal submatrix $(A_{i,j})_{0 \le i,j \le n}$, $l,u$ are $m \times m$ lower and upper triangular blocks, and $M$ and $V$ are rectangular, $m \times (n+1)$ and $(n+1) \times m$ respectively. 

It is not difficult to see that $M$ and $V$ are given by
\be
M_{a,j} = L_{n+k_a,j}, \qquad V_{i,b} = U_{i,n+\ell_b}, \qquad 1 \le a,b \le m, \: 0 \le i,j \le n.
\ee
One verifies for instance, for $j \le n$,
\be
(MU)_{a,j} = \sum_{i=0}^n \, L_{n+k_a,i} \, U_{i,j} = \sum_{i \ge 0} \, L_{n+k_a,i} \, U_{i,j} = A_{n+k_a,j},
\ee
and likewise for $LV$. From this, we find that
\be
g_{a,b} \equiv (lu)_{a,b} = A_{n+k_a,n+\ell_b}(\vec k|\vec \ell) - (MV)_{a,b} = \sum_{j=1}^{\min(k_a,\ell_b)} \, L_{n+k_a,n+j} \, U_{n+j,n+\ell_b}, \quad 1 \le a,b \le m,
\ee
is the matrix generalization of the quantity we called $g$ in the previous section. We thus obtain,
\be
\frac{Z_{n+m}(\vec k|\vec\ell)}{Z_n} = \det\big(Z_1^{\rm out}(k_a|\ell_b)\big)_{1 \le a,b \le m} = \det(g_{a,b})_{1 \le a,b \le m},
\ee
a formula that is surprisingly reminiscent of an LGV determinant (it is not really an LGV determinant since the numbers $g_{a,b}$ are not weighted sums over paths). Incidentally, and since all the entries of $L$ and $U$ are integers for $w=1$ (or any positive integer), the formula shows that in this case, all ratios $Z_{n+m}(\vec k|\vec\ell)/Z_n$ are integers, something that was not a priori obvious.

In the scaling limit, we set $k_a=r_an$ and $\ell_b=s_bn$, and using the results of the previous section for the asymptotic value of $g_{a,b}$, namely,
\be
g_{a,b} = f_{a,b} \, \exp\{n S(r_a,s_b)\}, \qquad \lim_{n \to \infty} \frac 1n \log f_{a,b} = 0,
\label{gab}
\ee
we have,
\be
\frac{Z_{n+m}(nr_1,\ldots,nr_m|ns_1,\ldots,ns_m)}{Z_n} = \det\big(f_{a,b} \, {\rm e}^{nS(r_a,s_b)}\big)_{1 \le a,b \le m}.
\label{Zkout}
\ee

Expanding the determinant as a sum over the symmetric group $S_m$, we obtain
\be
\lim_{n \to \infty} \frac 1n \log\frac{Z_{n+m}(nr_1,\ldots,nr_m|ns_1,\ldots,ns_m)}{Z_n} \le \max_{\pi \in S_m} \big\{ S(r_1,s_{\pi(1)}) + \ldots + S(r_m,s_{\pi(m)})\big\}.
\label{bound}
\ee
In case the maximum is attained for more than one permutation, subtle cancellations could occur among the factors $f_{a,b}$, resulting in a strict inequality, although this is not to be expected. We will show that (1) the identity permutation is always among the permutations for which the maximum is reached, though it is in general not the only one, and (2) that no cancellations occur if, for those $a$ such that $r_a s_a > \frac 1{1+w}$, there is no equality between two $r_a$ or between two $s_a$. The proofs of both claims rely on the property that the difference $S(r,s) - S(r,s')$, for any $s \ge s'$, is an increasing function of $r$; the proof is given in Appendix \ref{appB}.

For the statement (1), we first consider $m=2$. In this case, since $r_2 \ge r_1$, we have indeed, 
\be
[S(r_1,s_1) + S(r_2,s_2)] - [S(r_1,s_2) + S(r_2,s_1)] = [S(r_2,s_2) - S(r_2,s_1)] - [S(r_1,s_2) - S(r_1,s_1)] \ge 0.
\ee
For general $m$, we proceed by recurrence, assuming that the identity permutation in $S_{m-1}$ yields a maximum. Let $\pi \in S_m$ be a permutation such that $\pi(m) \neq m$ (otherwise we are done). Then
\bea
\sum_{a=1}^m \: [S(r_a,s_a) - S(r_a,s_{\pi(a)})] && \nonumber\\
&& \hspace{-4.7cm} = \, [S(r_m,s_m) - S(r_m,s_{\pi(m)})] - [S(r_{\pi^{-1}(m)},s_m) - S(r_{\pi^{-1}(m)},s_{\pi^{-1}(m)})] + \ldots \nonumber\\
\noalign{\medskip}
&& \hspace{-4.7cm} \ge \, [S(r_{\pi^{-1}(m)},s_m) - S(r_{\pi^{-1}(m)},s_{\pi(m)})] - [S(r_{\pi^{-1}(m)},s_m) - S(r_{\pi^{-1}(m)},s_{\pi^{-1}(m)})] + \ldots \nonumber\\
\noalign{\medskip}
&& \hspace{-4.7cm} = \, [S(r_{\pi^{-1}(m)},s_{\pi^{-1}(m)}) - S(r_{\pi^{-1}(m)},s_{\pi(m)})]  + \sum_{a \neq m,\pi^{-1}(m)} \: [S(r_a,s_a) - S(r_a,s_{\pi(a)})] \nonumber\\
&& \hspace{-4.7cm} = \, \sum_{a=1}^{m-1} \: [S(r_a,s_a) - S(r_a,s_{\tilde\pi(a)})] \ge 0,
\label{ineq}
\eea
where $\tilde\pi \in S_{m-1}$ is the permutation acting on $\{1,2,\ldots,m-1\}$ as
\be
\tilde\pi(a) = \begin{cases}
\pi(a) & {\rm for\ } a \neq \pi^{-1}(m),\\
\pi(m) & {\rm for\ } a = \pi^{-1}(m).
\end{cases}
\ee  
As the claim is true for $m=2$, it holds for every $m$.

Let $m_0$ be such $r_1 s_1 \le \ldots \le r_{m_0}s_{m_0}\le \frac 1{1+w} < r_{m_0+1}s_{m_0+1} \le \ldots \le r_ms_m$. For the point (2), we first characterize, under the assumptions stated (no two $r$ and no two $s$ with labels strictly larger than $m_0$ are equal), the permutations for which the maximum is attained, namely those $\pi$ such that 
\be
\sum_{a=1}^m S(r_a,s_{\pi(a)}) = \sum_{a=1}^m S(r_a,s_a).
\label{sum}
\ee 
%Because the two finite sequences $(r_a)_{1 \le a \le m}$ and $(s_a)_{1 \le a \le m}$ are (separately) ordered, there is unique integer $m_0$, between 0 and $m$, such that $r_a s_a \le \frac 1{1+w}$ for all $a \le m_0$ and $r_a s_a > \frac 1{1+w}$ for all $a \ge m_0+1$.
We note that the group $S_{m_0}$ of permutations of the labels $a \le m_0$ satisfy this identity since $S(r,s)$ has the form $\sigma(r) + \sigma(s)$ for such pairs $(r,s)$. We prove that (\ref{sum}) is satisfied for no other permutation.

Let $\pi$ be such a non-trivial permutation, so that there exists a label $b$ with $\pi(b) \neq b$ such that either $b > m_0$ or $\pi(b) > m_0$, or both. Without loss of generality, we may assume that $b > m_0$ and also $s_{b} > s_{\pi(b)}$, $r_{b} > r_{\pi^{-1}(b)}$. In this case, we repeat the chain of (in)equalities in (\ref{ineq}) in which we replace, in the second line, the label $m$ by $b$.  However on the third line, we obtain a strict inequality since with the conditions we have just stated, the difference $S(r_{b},s_{b}) - S(r_{b},s_{\pi(b)})$ is strictly larger than $S(r_{\pi^{-1}(b)},s_{b}) - S(r_{\pi^{-1}(b)},s_{\pi(b)})$, as shown in Appendix \ref{appB}.

Coming back to the determinant (\ref{Zkout}), we see that the terms having the same exponential rate in $n$ are given by
\bea
\frac{Z_{n+m}(nr_1,\ldots,nr_m|ns_1,\ldots,ns_m)}{Z_n} &\!\!\!\simeq\!\!\!& \det\big(f_{a,b} \, {\rm e}^{n\sigma(r_a)+n\sigma(s_b)}\big)_{1 \le a,b \le m_0} \times \Big[\prod_{a=m_0+1}^m f_{a,a} \, {\rm e}^{nS(r_a,s_a)}\Big] \nonumber\\
&&\hspace{-2cm} = \: \det\big(f_{a,b}\big)_{1 \le a,b \le m_0} \times \Big[\prod_{a=m_0+1}^m f_{a,a}\Big] \times \exp\Big\{n\sum_{a=1}^{m} S(r_a,s_a)\Big\}.
\eea
From the definition (\ref{gab}), none of the coefficient $f_{a,a}$ vanishes. The determinant does not vanish either because the partition function for the $m_0$ extra paths alone is proportional to it,
\bea
\frac{Z_{n+m_0}(nr_1,\ldots,nr_{m_0}|ns_1,\ldots,ns_{m_0})}{Z_n} \egal \det\big(f_{a,b} \, {\rm e}^{nS(r_a,s_b)}\big)_{1 \le a,b \le m_0} \nonumber\\
\egal \det\big(f_{a,b}\big)_{1 \le a,b \le m_0} \times \exp\Big\{n\sum_{a=1}^{m_0} S(r_a,s_a)\Big\}.
\eea

Under the assumptions we made, namely $r_a \neq r_b$ and $s_a \neq s_b$ for $a,b \ge m_0+1$, we have thus proved that the upper bound is (\ref{bound}) is reached. The completely general situation corresponds to the existence of subsets $T_i$ of labels larger than $m_0$ yielding identical $r$-values or $s$-values. Using the same arguments as above, we can extend the proof  provided the subsets $T_i$ are disjoint. Whether the $T_i$ subsets are disjoint or not, the permutations satisfying (\ref{sum}) can be determined explicitly. However when the $T_i$ are not disjoint\footnote{The simplest situation of that type involves three doublets $(r_1,s_1)$, $(r_2,s_2)$ and $(r_3,s_3)$ with $r_1=r_2$ and $s_2=s_3$, corresponding to $T_1=\{1,2\}$ and $T_2=\{2,3\}$. In this case, the permutations which satisfy (\ref{sum}) are $\pi_1(1,2,3)=(2,1,3)$, $\pi_2(1,2,3)=(1,3,2)$ and the identity. They do not form a subgroup of $S_3$.}, these permutations do not form a direct product of permutations subgroups (they do not even form a group), and therefore do not build full subdeterminants (as the group $S_{m_0}$ above did). The above arguments then fail. But as these special cases  are limits of cases for which our proof works, the result presumably holds for them as well, by continuity.

To summarize, we have shown that the following equality generically holds, 
\be
\lim_{n \to \infty} \frac 1n \log\frac{Z_{n+m}(nr_1,\ldots,nr_m|ns_1,\ldots,ns_m)}{Z_n} = S(r_1,s_1) + S(r_2,s_2) + \ldots + S(r_m,s_m),
\ee
thereby proving the general conjecture in the case of Aztec diamonds.

\section{Multirefinements for alternating sign matrices}
\label{sec4}

In this section we show that the factorization hypothesis appears to apply to non-crossing paths that may have kissing points. We will verify it in the case of multirefined enumerations of alternating sign matrices (ASM). 

An ASM of size $n$ is an $n\times n$ matrix with entries 1, 0, and $-1$ such that the 1 and $-1$ entries alternate in every column and every row and such that all row and column sums are equal to 1 (first and last non-zero entries are 1's). We will denote by ASM$_n$ the set of ASM of order $n$, and mainly consider the uniform distribution on ASM$_n$. This model is equivalent to the 6-vertex model with domain wall boundary conditions and parameters $\Delta=\frac 12$ and $t=1$.

%{\red More generally one can also consider the six-vertex model with domain wall boundary conditions, which is equivalent to the ASM enumeration when its parameters are specialized to $\Delta=\frac 12$ and $t=1$. Appendix \ref{appA} shows how to compute the L function in this case.}

%, also called the 1-enumeration, equivalently the 6-vertex model at $\Delta=\frac 12$ and $t=1$. In the $q$-enumeration, each ASM is weighted by the factor $q^k$ with $k$ the number of $-1$ the matrix contains. The case $q=2$, corresponding to the 6-vertex model with $\Delta=0$, is equivalent to the tiling of Aztec diamonds discussed in Section \ref{sec3}, although in terms of a different path description. For $q=3$, equivalently the 6-vertex model at $\Delta=-\frac 12$, exact 1-refined partition functions are known for $q=3$ \cite{CP05}, as is the arctic curve \cite{CP10}. However the nature of the arctic curve, an algebraic curve of degree 6, leaves little hope that the function $L$ computed in Appendix \ref{appA} can be integrated analytically. 

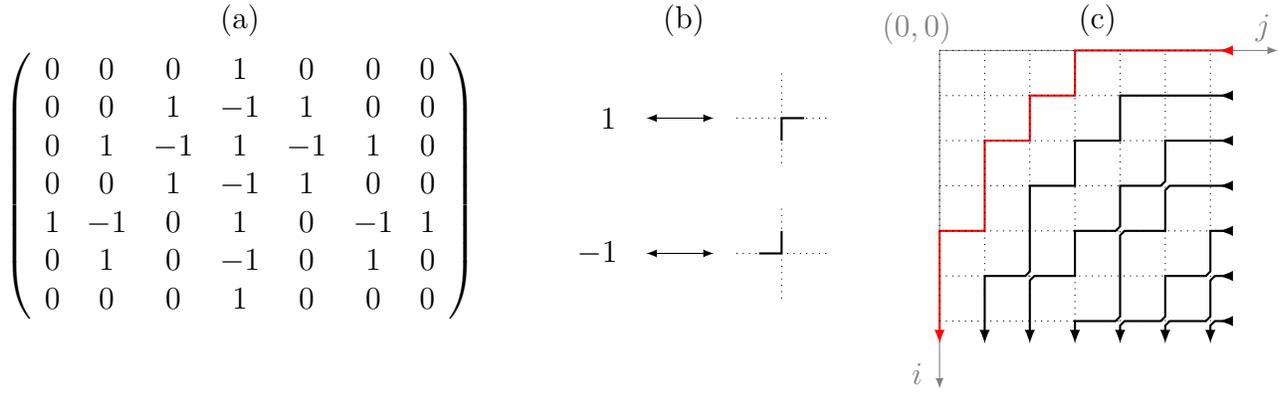
\begin{figure}[t]
\begin{center}
\begin{tikzpicture}
\draw (-3.2+3.2,2.2) node{\text{(a)}};
\draw (-3.2+7.5+1.6,2.2) node{\text{(b)}};
\draw (-3.2+12.5+2.1,2.2) node{\text{(c)}};
\draw (0,0) node{$\left(
\begin{array}{ccccccc}
0 & 0 & 0 & 1 & 0 & 0 & 0 \\ 
0 & 0 & 1 & -1 & 1 & 0 & 0 \\ 
0 & 1 & -1 & 1 & -1 & 1 & 0 \\ 
0 & 0 & 1 & -1 & 1 & 0 & 0 \\ 
1 & -1 & 0 & 1 & 0 & -1 & 1 \\ 
0 & 1 & 0 & -1 & 0 & 1 & 0 \\ 
0 & 0 & 0 & 1 & 0 & 0 & 0
\end{array} 
\right)$};
\begin{scope}[scale=0.6,xshift=15.5cm, yshift=-3cm]

\draw [gray,<->, >=latex] (0,-1.5)node[xshift=-0.3cm,yshift=0.2cm]{$i$}--(0,6) node[xshift=-0.3cm,yshift=0.3cm]{$(0,0)$}--(7.5,6)node[xshift=-0.2cm,yshift=0.3cm]{$j$};
\draw[<-<,>=latex,red,thick] (0,-0.5)--++(0,0.5)--++(0,2)--++(1,0)--++(0,2)--++(1,0)--++(0,1)--++(1,0)--++(0,1)--++(3,0)--++(0.5,0);
\draw[<-<,>=latex,thick] (1,-0.5)--++(0,0.5)--++(0,1)--++(0.9,0)--++(0.1,0.1)--++(0,1.9)--++(1,0)--++(0,1)--++(1,0)--++(0,1)--++(2,0)--++(0.5,0);
\draw[<-<,>=latex,thick] (2,-0.5)--++(0,0.5)--++(0,0.9)--++(0.1,0.1)--++(0.9,0)--++(0,1)--++(0.9,0)--++(0.1,0.1)--++(0,0.9)--++(0.9,0)--++(0.1,0.1)--++(0,0.9)--++(1,0)--++(0.5,0);
\draw[<-<,>=latex,thick] (3,-0.5)--++(0,0.5)--++(0.9,0)--++(0.1,0.1)--++(0,1.8)--++(0.1,0.1)--++(0.9,0)--++(0,0.9)--++(0.1,0.1)--++(0.9,0)--++(0.5,0);
\draw[<-<,>=latex,thick] (4,-0.5)--++(0,0.4)--++(0.1,0.1)--++(0.8,0)--++(0.1,0.1)--++(0,0.9)--++(0.9,0)--++(0.1,0.1)--++(0,0.9)--++(0.5,0);
\draw[<-<,>=latex,thick] (5,-0.5)--++(0,0.4)--++(0.1,0.1)--++(0.8,0)--++(0.1,0.1)--++(0,0.8)--++(0.1,0.1)--++(0.4,0);
\draw[<-<,>=latex,thick] (6,-0.5)--++(0,0.4)--++(0.1,0.1)--++(0.4,0);
\foreach \i in {0,1,...,6}{
\draw[dotted](0,\i)--++(6,0);
\draw[dotted](\i,0)--++(0,6);}
%\draw (-0.25,0)--++(0.25,0)--++(3,0)--++(0,1)--++(2,0)--++(0,1)--++(1,0)--++(0,4)--++(0,0.25);
%\draw (-0.25,1)--++(0.25,0)--++(1,0)--++(0,1)--++(2,0)--++(0,1)--++(1,0)--++(0,1)--++(1,0)--++(0,2)--++(0,0.25);
%\draw (-0.25,2)--++(0.25,0)--++(0,0.9)--++(0.1,0.1)--++(2,0)--++(0,1)--++(1,0)--++(0,1)--++(1,0)--++(0,1)--++(0,0.25);
%\draw (-0.25,3)--++(0.15,0)--++(0.1,0.1)--++(0,0.8)--++(0.1,0.1)--++(1,0)--++(0,0.9)--++(0.1,0.1)--++(0.9,0)--++(0,0.9)--++(0.1,0.1)--++(0.9,0)--++(0,0.25);
%\draw (-0.25,4)--++(0.15,0)--++(0.1,0.1)--++(0,0.8)--++(0.1,0.1)--++(0.8,0)--++(0.1,0.1)--++(0,0.8)--++(0.1,0.1)--++(0.8,0)--++(0.1,0.1)--++(0,0.15);
%\draw (-0.25,5)--++(0.15,0)--++(0.1,0.1)--++(0,0.8)--++(0.1,0.1)--++(0.8,0)--++(0.1,0.1)--++(0,0.15);
%\draw (-0.25,6)--++(0.15,0)--++(0.1,0.1)--++(0,0.15);
\end{scope}
\begin{scope}[scale=0.6,xshift=9cm, yshift=+1.5cm]
\draw[<->,>=latex] (0,0) node[xshift=-0.5cm]{$1$} -- (1.5,0);
\draw[dotted] (2,0)--++(2,0);
\draw[dotted] (3,-1)--++(0,2);
\draw[thick] (3,-0.5)--++(0,0.5)--++(0.5,0);
\begin{scope}[yshift=-3cm]
\draw[<->,>=latex] (0,0) node[xshift=-0.65cm]{$-1$} -- (1.5,0);
\draw[dotted] (2,0)--++(2,0);
\draw[dotted] (3,-1)--++(0,2);
\draw[thick] (2.5,0)--++(0.5,0)--++(0,0.5);
\end{scope}
\end{scope}
\end{tikzpicture}
\end{center}
\vspace{-0.5cm}
\caption{\textbf{(a)} Example of ASM of order $n=7$. The construction of the corresponding non-crossing paths presented in \textbf{(c)} is done by replacing the 1 and $-1$ by portions of paths as shown in \textbf{(b)} and by completing the remaining portions in a way to obtain $n$ paths that start from the right boundary and arrive at the bottom boundary  without crossing. In the scaling limit, the outermost path, in red, condenses onto a portion of arctic curve.}
\label{fig_ASM_non_crossing}
\end{figure}

A matrix in ASM$_n$ can be represented bijectively as a set of $n$ non-crossing (osculating) lattice paths drawn on an $n \times n$ grid. The presence of a $1$ or a $-1$ in the matrix at position $(i,j)$ means that the site $(i,j)$ is visited by strictly one of the $n$ paths, which in addition makes at that site one of the two turns shown on Figure \ref{fig_ASM_non_crossing}. These fix completely the rest of the $n$ paths, required to connect the $n$ vertices of the last column and those on the last row without crossing each other, although they are allowed to touch (they can share sites but not edges). If we see the entries of the last column as entrance points and those on the last row as exit points, then, in the $(i,j)$ plane, the paths are made of the elementary steps $(1,0)$ and $(0,-1)$. These elementary steps are respectively called ``right'' and ``down'', when viewed in the reference frame of Figure \ref{fig_ASM_non_crossing}(c).

The path representation allows to apply the tangent method but prevents the use of the LGV lemma to compute the partition function, as the paths in general are not non-intersecting. The expression for the partition function for ASM$_n$ was proved in \cite{Ze96a,Ku96},
\be
A_n = \prod_{j=0}^{n-1} \:\frac{(3j+1)!}{(n+j)!} = \prod_{j=0}^{n-1} \:\frac{{3j+1 \choose j}}{{2j \choose j}}.
\label{asm}
\ee

\begin{figure}[t]
\begin{center}
\includegraphics[scale=1]{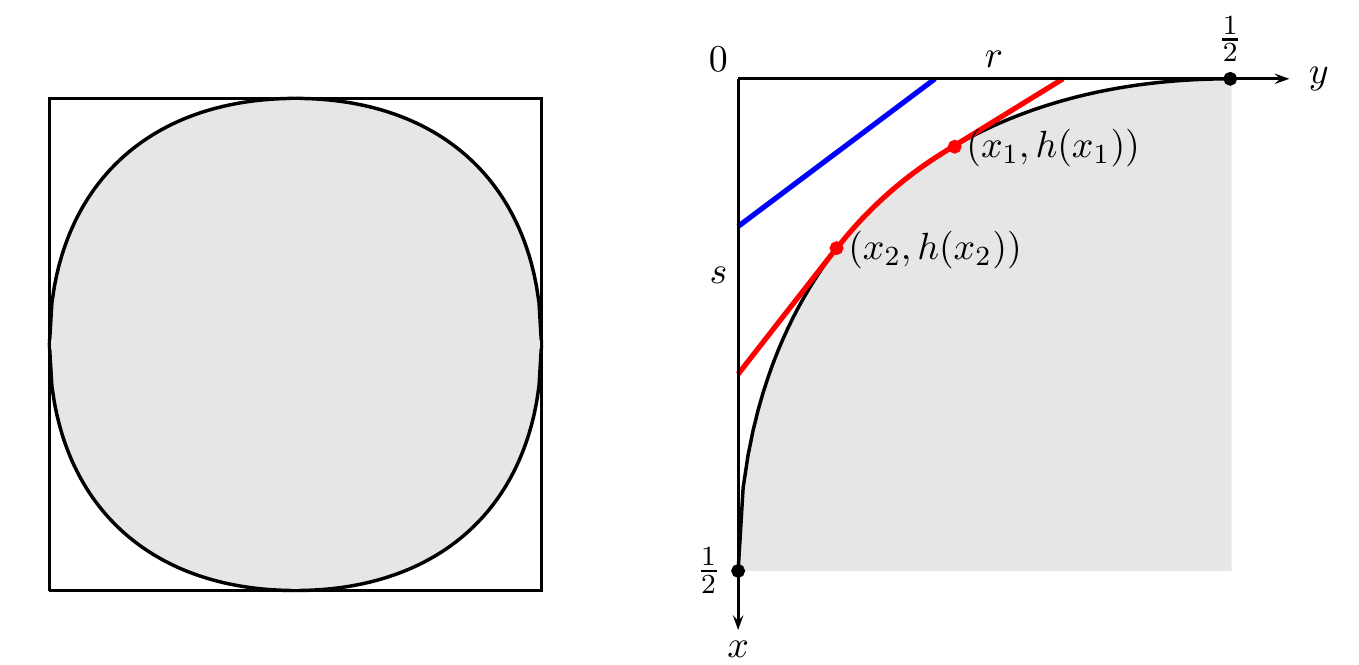}
\end{center}
\vspace{-1cm}
\caption{On the left, the picture shows the arctic curve for the ASM, with the disordered region shaded. The right panel shows the NW quadrant magnified and the maximizing trajectory $f^*$ in the two typical cases of 2-refined ASM in the scaling limit: $f^*$ touches or does not touch the arctic curve depending on the sign of $2r+2s-rs-1$.}
\label{fig_AC_ASM}
\end{figure}

In the scaling limit, ASM exhibit an arctic curve which separates the central region where the three entries $0,-1$ and $+1$ are present with non-zero densities, from four disconnected regions around the corners, which are filled with $0$'s with probability 1, see Figure \ref{fig_AC_ASM}. The exact shape of the arctic curve was first conjectured in \cite{CP10a} and rigorously computed in \cite{Ag20}. In the coordinate system used in Figure \ref{fig_AC_ASM}, in which $x$ and $y$ are the continuous counterparts of the row and column labels, the northwestern (NW) portion of the arctic curve satisfies the conic equation $x(1-x) + y(1-y) + xy = \frac 14$, for $0 \le x,y \le \frac 12$, from which the upper branch is selected,
\be 
h(x)=\frac{1}{2}\big(1 + x - \sqrt{3x(2-x)} \big), \qquad 0 \le x \le \frac 12.
\ee
The other three portions can be deduced from mirror symmetry with respect to the horizontal and the vertical central axes. In what follows we focus on the NW portion. Other portions may also be studied by using other path descriptions.

An easy property of ASM is that they all have a single 1 (thus no $-1$) on the first and last rows and columns. For large and generic ASM, each of these 1's are almost surely located in the scaling neighbourhood of the middle point of their row or column. In the discrete setting, this means that the uppermost path takes its first step rightwards, i.e. $(1,0)$, around the position $(i,j)=(1,\frac n2)$ and its last step downwards around the position $(\frac n2,2)$; in between these two positions it follows a zig-zag trajectory around the arctic curve. In the scaling limit and in the reference frame of Figure \ref{fig_AC_ASM}, the uppermost path starts from $(0,1)$, joins the point $(0,\frac 12)$ vertically, follows the arctic curve till $(\frac 12,0)$, and finally moves horizontally to reach the point $(1,0)$.

Two-refined partition functions can then be defined by considering ASM with a $1$ at fixed positions on the first row and first column, $m_{1,k}=1$ and $m_{\ell,1}=1$, with $k,\ell \le \frac n2$. For these, the uppermost path leaves the first row at position $(1,k)$, and enters the first column at position $(\ell,1)$. In the scaling limit, depending on the values of $k=rn$ and $\ell=sn$, the uppermost path will follow a straight line from $(0,r)$ to $(s,0)$, or will follow a portion of the arctic curve, as illustrated in Figure \ref{fig_AC_ASM}. The unrefined case is recovered for $r=s=\frac12$. 

If we denote by $A_{n+1}(k|\ell)$ the number of such 2-refined ASM of order $n+1$, the factorization conjecture implies, 
\be
S(r,s) \equiv \lim_{n \to \infty} \frac 1n \log \frac{A_{n+1}(rn|sn)}{A_n} = S[f^*] = \int_{0}^{1} {\rm d}x \, L\Big(\frac{{\rm d}f^*}{{\rm d}x}\Big), \qquad 0 \le r,s \le \frac 12,
\ee
where $f^*$ is the trajectory followed by the uppermost path in the scaling limit: it goes from $f^*(0)=1$ to $f^*(1)=0$ by passing through $f^*(0)=r, \, f^*(s)=0$ and minimizes the distance without ever crossing the arctic curve. The relevant function $L$ is the one that pertains to the paths used for the ASM.

For multirefined partition functions, we would consider the number $A_{n+m}(k_1,\ldots,k_m|\ell_1,\ldots,\ell_m)$ of ASM such that in the scaling limit, their $j$-th uppermost path leaves the top boundary at $(0,r_j)$ and reaches the left boundary at $(s_j,0)$, with $0 \le r_1 \le r_2 \le \ldots \le r_m \le \frac 12$ and $0 \le s_1 \le s_2 \le \ldots \le s_m \le \frac 12$. Then a formula similar to (\ref{mref}) is conjectured to hold, where the actions $S(r_i,s_i)$ for the individual paths are simply added.

%%%%%%%%%%%%%%%%%%%%%%%%%%%%%%%%%%%%%%%%%
\subsection{The 2-refined ASM entropy function}

The paths that make up the correspondence with ASM are composed of two elementary steps $(1,0)$ and $(0,-1)$, both with weight 1. The appropriate Lagrangean function then reads \cite{DGR19}
\be 
L(t) = (1 - t) \log(1 - t) + t \log |t|, \qquad t \le 0.
\ee
It satisfies $L(0)=0$ and $\lim_{t \to -\infty} L(t) = 0$. The function $L$ for the general 6-vertex model is given in Appendix \ref{appA}.

To compute the function $S(r,s)$, we first note that the horizontal and vertical portions of $f^*$ do not contribute to the integral since $L(0) = L(-\infty) = 0$ as we have just noted.  If $r$ and $s$ are close enough to $\frac 12$, the trajectory $f^*$ is rectilinear from $(0,r)$ to a first tangency point $\big(x_1(r),h(x_1(r))\big)$, then follows the arctic curve till a second tangency point $\big(x_2(s),h(x_2(s))\big)$, where it tangentially leaves the arctic curve and moves rectilinearly to the endpoint $(s,0)$. Fixing $r$ and $s$, one finds, for $t_i = h'(x_i)$,
\be
t_1=-\frac{r(2-r)}{1-2r}, \quad x_1 = 1-\frac{1-2t_1}{2\sqrt{1-t_1+t_1^2}}, \qquad t_2 = -\frac{1-2s}{s(2-s)}, \quad x_2 = 1-\frac{1-2t_2}{2\sqrt{1-t_2+t_2^2}},
\ee
where $t_1 < t_2$ are both negative. As $r$ and $s$ decrease, the slopes $t_1$ and $t_2$ respectively increases and decreases, until they become equal, implying also that $x_1=x_2$. If $r,s$ decrease even further, the trajectory $f^*$ is a single straight line going from $(0,r)$ to $(s,0)$. The borderline between the two cases corresponds to $t_1(r) = t_2(s)$, or $2r+2s-rs=1$, which, in the $(r,s)$-plane, is a concave curve slightly above the antidiagonal $2r+2s=1$.

When $r,s$ are such that $t_1(r) \le t_2(s)$, we have 
\be
S(r,s) = S[f^*] = x_1 \, L(t_1) + \int_{x_1}^{x_2} {\rm d}x \; L\big(h'(x) \big) + (s-x_2) \, L(t_2),
\ee
which, like the similar calculation in the Aztec diamonds, reduces to a simple symmetric function of $r,s$. For $t_1(r) \ge t_2(s)$, $f^*$ is a straight line so that the integral is straightforward.

The final results read, for $0 \le r,s \le \frac 12$,
\be
S(r,s) = \begin{cases}
\displaystyle s \, L(\textstyle -\frac rs) = (r+s)\log(r+s) - r\log r - s\log s, & \hspace{-4mm} {\rm if\ } 2r+2s-rs \le 1, \\
\noalign{\medskip}
(1+r) \log(1+r) - r \log r + (2-r) \log(2-r) - (1-r) \log(1-r)  & \\
\hspace{2truecm} + \: (r \to s) - 3 \log 3, & \hspace{-4mm} {\rm if\ } 2r+2s-rs \ge 1.
\end{cases}
\label{Srs2}
\ee

As a first simple check, we consider the unrefined partition functions for ASM of order $n+1$ and order $n$. Since the unrefined case corresponds to $r=s=\frac 12$, the factorization implies
\be
\lim_{n \to \infty} \frac 1n \log \frac{A_{n+1}}{A_n} = \lim_{n \to \infty} \frac 1n \log \frac{A_{n+1}(\frac n2|\frac n2)}{A_n} = \textstyle S(\frac 12,\frac 12) = \log \frac{27}{16},
\ee
where we have used the second form in (\ref{Srs2}). From (\ref{asm}), the exact ratio is equal to $A_{n+1}/A_n = {3n+1 \choose n}/{2n \choose n}$, and is asymptotic to $\big(\frac {27}{16}\big)^n$, in agreement with the previous value.

Another limiting case is when the refined matrices in ASM$_{n+1}$ have $m_{1,1}=1$, in such a way that this 1 is shared between the first row and the first column. It corresponds to $k=\ell=1$, and $r=s=0$ in the scaling limit. As the number of such refined matrices in ASM$_{n+1}$ is equal to the unrefined matrices in ASM$_n$, the ratio $A_{n+1}(1|1)/A_n$ is exactly equal to 1, implying $S(0,0)=0$. This is also the value obtained from the first form of (\ref{Srs2}).

%%%%%%%%%%%%%%%%%%%%%%%%%%%%%%%%%%%%%%%%%
\subsection{Refined alternating sign matrices}

In this section we study a selection of multirefined enumerations of ASM. A quite extended literature exists on the subject, see for instance \cite{Be13}. The results recalled here are collected in \cite{AR13}. 

The simplest case is that of the 1-refined enumeration of ASM, by which one counts the number $A_n(k|-)$ of matrices in ASM$_n$ having their unique 1 in the first row at position $k$, without any condition on the unique 1 in the first column. The result is surprisingly simple \cite{Ze96b}
\be 
A_n(k|-) = {n+k-2 \choose k-1} \frac{(2n-k-1)!}{(n-k)!} \: \prod_{j=0}^{n-2} \frac{(3j+1)!}{(n+j)!}.
\ee
It follows that the effective contribution of the uppermost path is given by
\be 
Z^\text{out}_1(k|-) = \frac{A_{n+1}(k|-)}{A_n} = \frac{{n+k-1 \choose n}{2n-k+1 \choose n}}{{2n \choose n}}.
\ee
from which, upon setting $k=rn$ with $r \le \frac 12$, one readily obtains
\bea
\lim_{n\to\infty} \frac{1}{n} \log Z^\text{out}_1(rn|-) \egal (1+r) \log(1+r) - r \log r + (2-r) \log(2-r) , \nonumber\\
&& - \: (1-r) \log(1-r) - 2 \log 2.
\label{1asm}
\eea
As no condition on the rescaled position $s=\ell/n$ of the unique 1 in the first column implies $s=\frac12$ almost surely, the previous limit is equal to $\lim_{n\to\infty} \frac{1}{n} \log Z^\text{out}_1(rn|\frac n2)$, expected to be equal to the function $S(r,\frac 12)$ computed in (\ref{Srs2}), which it is. 

\begin{figure}[t]
\begin{tikzpicture}
%\draw (0,0) node{\includegraphics[scale=0.35]{Two_TB_refined_ASM_Zup_n400_exact.eps}};
%\draw(0,3.5) node{$\frac{1}{n} \log(A^\text{TB}_n(rn,sn))$};
%\draw (0,0) node{\includegraphics[scale=0.35]{Two_TB_refined_ASM_Zup_n400_factorization.eps}};
\draw (0,0) node{\includegraphics[scale=0.35]{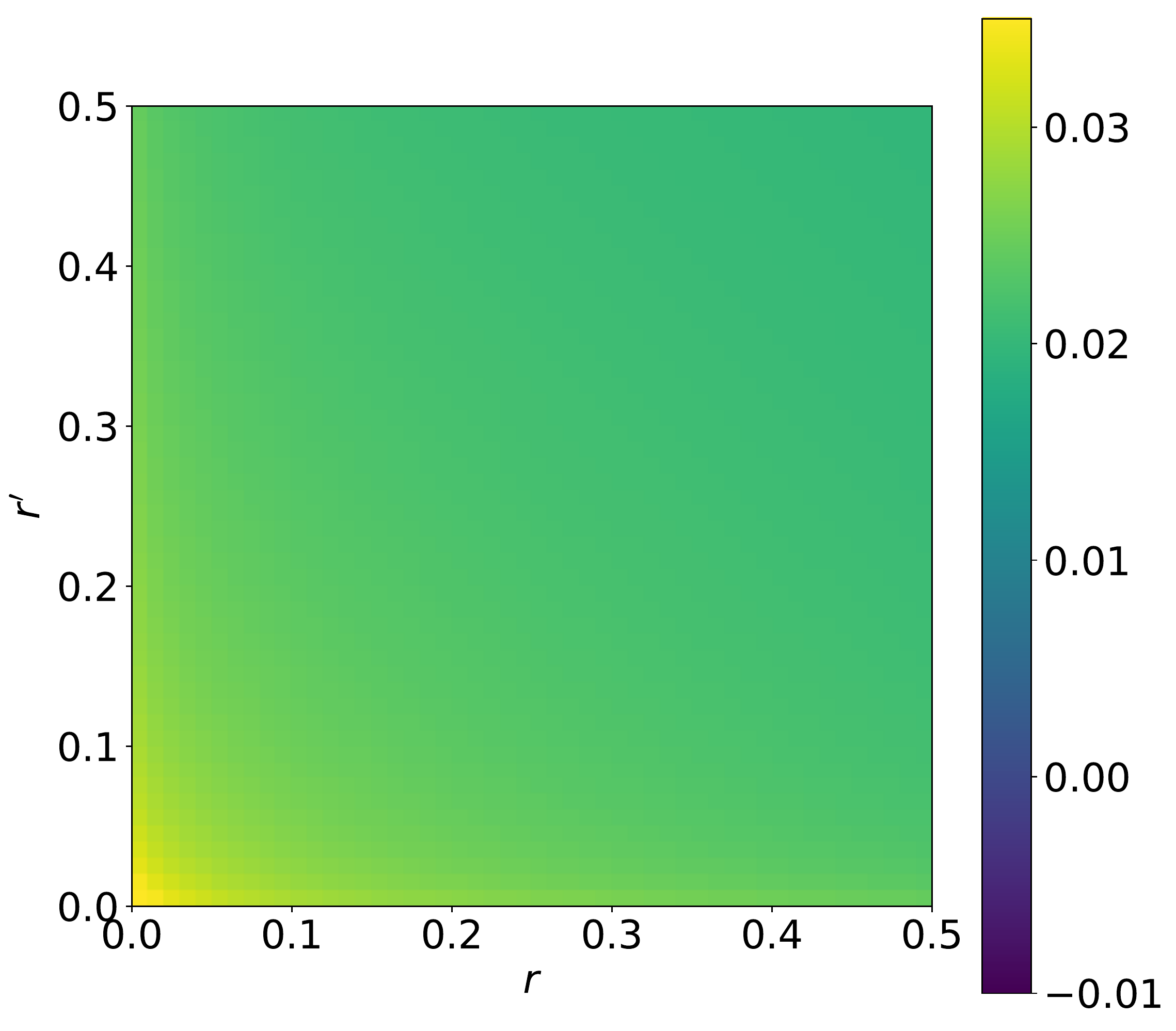}};
\draw(-.7,3.5) node{\small $S(r,\frac 12)+S(r',\frac 12)-\frac{1}{n} \log\frac{A^\text{TB}_{n+2}(rn,r'n)}{A_n}$};
\draw (9.3,0) node{\includegraphics[scale=0.35]{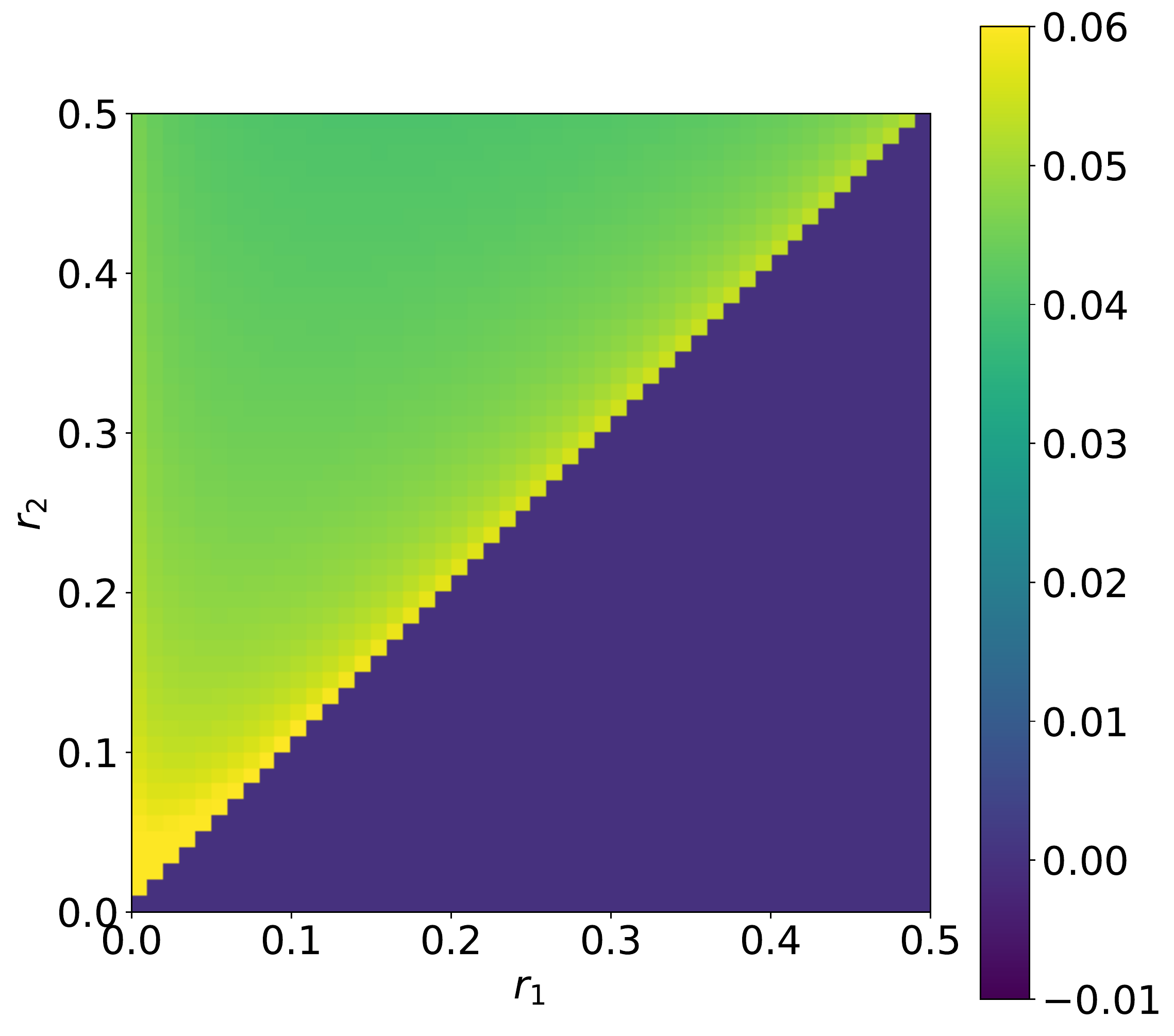}};
\draw(8.5,3.5) node{\small $S(r_1,\frac 12)+S(r_2,\frac 12)-\frac{1}{n} \log\frac{A^\text{TT}_{n+2}(r_1n,r_2n)}{A_n}$};
\end{tikzpicture}
\caption{Verification of the factorization hypothesis for the TB (left) and TT (right) refinements of ASM for $n=400$ and $n=200$ respectively.}
\label{fig_TB_TT}
\end{figure}

There are two other refinements of ASM which are related to the 1-refined partition functions we have just discussed.
The first one is the so-called the Top-Bottom 2-refined enumeration $A_n^\text{TB}(k,k')$, which is the number of matrices in ASM$_n$ with their unique 1 on the top and bottom rows respectively in columns $k$ and $k'$. Strictly speaking, one cannot handle it within a single path description. However, because the two refinements concern regions of the domain which are far away, one would expect that they are weakly correlated, in such way that the following factorization would hold,
\be
\frac{A_{n+2}^\text{TB}(k,k')}{A_n} = \frac{A_{n+2}^\text{TB}(k,k')}{A_{n+1}^\text{B}(k')} \: \frac{A_{n+1}^\text{B}(k')}{A_n} \simeq \frac{A_{n+2}^\text{T}(k)}{A_{n+1}} \: \frac{A_{n+1}^\text{B}(k')}{A_n}.
\ee
%For $1 \le k \le k' \le n$ it is given by (\textcolor{red}{notice the typo in \cite{AR13}})
%\be 
%\begin{split}
%&A_n^\text{TB}(k,k')=A_{n-1,k'-k}+\sum_{\ell=1}^{k-1} D_n(\ell, k'-k+\ell),\\
%&D_n(s,t)=\frac{1}{A_{n-1}}(A_{n-1,t}(A_{n,s+1}-A_{n,s})+A_{n-1,s}(A_{n,t+1}-A_{n,t})).
%\end{split}
%\ee
In the scaling limit, and for $k=rn$ and $k'=r'n$ with $r,r' \le \frac 12$, it leads to two independent 1-refined contributions,
\be
\lim_{n \to \infty} \frac{1}{n} \log \frac{A_{n+2}^\text{TB}(rn,r'n)}{A_n} = S(r,\textstyle \frac 12) + S(r',\textstyle \frac 12).
\ee
The exact evaluation of $A_n^\text{TB}(k,k')$ was carried out in \cite{CP05,St06}, and rewritten in \cite{AR13} but the result does not easily lend itself to an asymptotic analysis. The previous equality can nonetheless be checked numerically for large $n$. The comparison is presented in the left panel of Figure \ref{fig_TB_TT}, which shows a good agreement\footnote{We have used the exact expressions from \cite{AR13}, after correcting a typo in their equation (1.3), where the first term on the right-hand side should read $A_{n-1,j-i}$ instead of $A_{n,j-i}$.}.

The second refinement is called Top-Top, and is based on the fact that there is a unique pair of column labels, $k_1 < k_2$, such that $m_{1,k_1} + m_{2,k_1} = m_{1,k_2} + m_{2,k_2} = 1$ \cite{FR09}. Let us denote by $A_{n+2}^{\rm TT}(k_1,k_2)$ the corresponding refined partition function. In terms of non-crossing paths, it means that the most upper path makes its first or second step rightwards at column $k_1$, and that the second most upper path makes its first step leftwards at column $k_2$. For large $n$, the partition function $A_{n+2}^{\rm TT}(k_1,k_2)$ can be assimilated to $A_{n+2}(k_1,k_2|-,-)$ defined earlier. For $k_1=r_1n$ and $k_2=r_2n$ with $r_1 \le r_2 \le \frac 12$, the factorization hypothesis predicts the same result as for the Top-Bottom case,
\be 
\lim_{n \to \infty} \frac{1}{n} \log\frac{A_{n+2}^{\rm TT}(r_1n,r_2n)}{A_{n}} = \lim_{n \to \infty} \frac{1}{n} \log\frac{A_{n+2}(r_1n,r_2n|-,-)}{A_{n}} = S(r_1,\textstyle\frac 12) + S(r_2,\frac 12).
\ee
The known determinations of $A_{n+2}^{\rm TT}(k_1,k_2)$ \cite{KR10,Fi11} do not allow to easily extract its asymptotics, so that we again relied on numerical evaluations. The right panel of Figure \ref{fig_TB_TT} confirms the validity of this identity.

\begin{figure}[t]
\begin{center}
\begin{tikzpicture}
%\draw (0,0) node{\includegraphics[scale=0.35]{Two_TT_refined_ASM_Zup_n200_exact.eps}};
%\draw(0,3.5) node{$\frac{1}{n} \log(A^\text{TT}_n(rn,sn))$};
\draw (10,0) node{\includegraphics[scale=0.35]{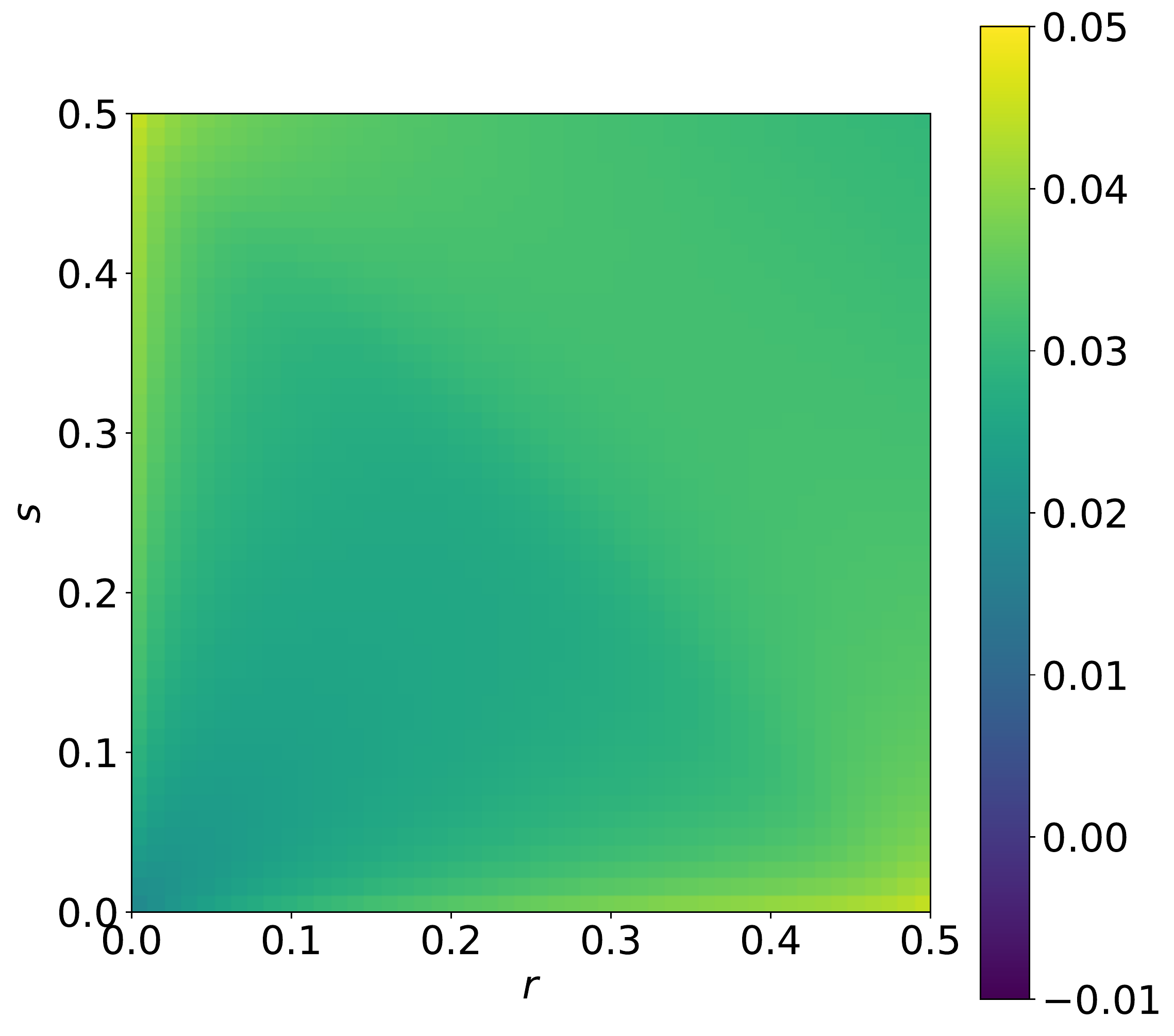}};
\draw (0,0) node{\includegraphics[scale=0.35]{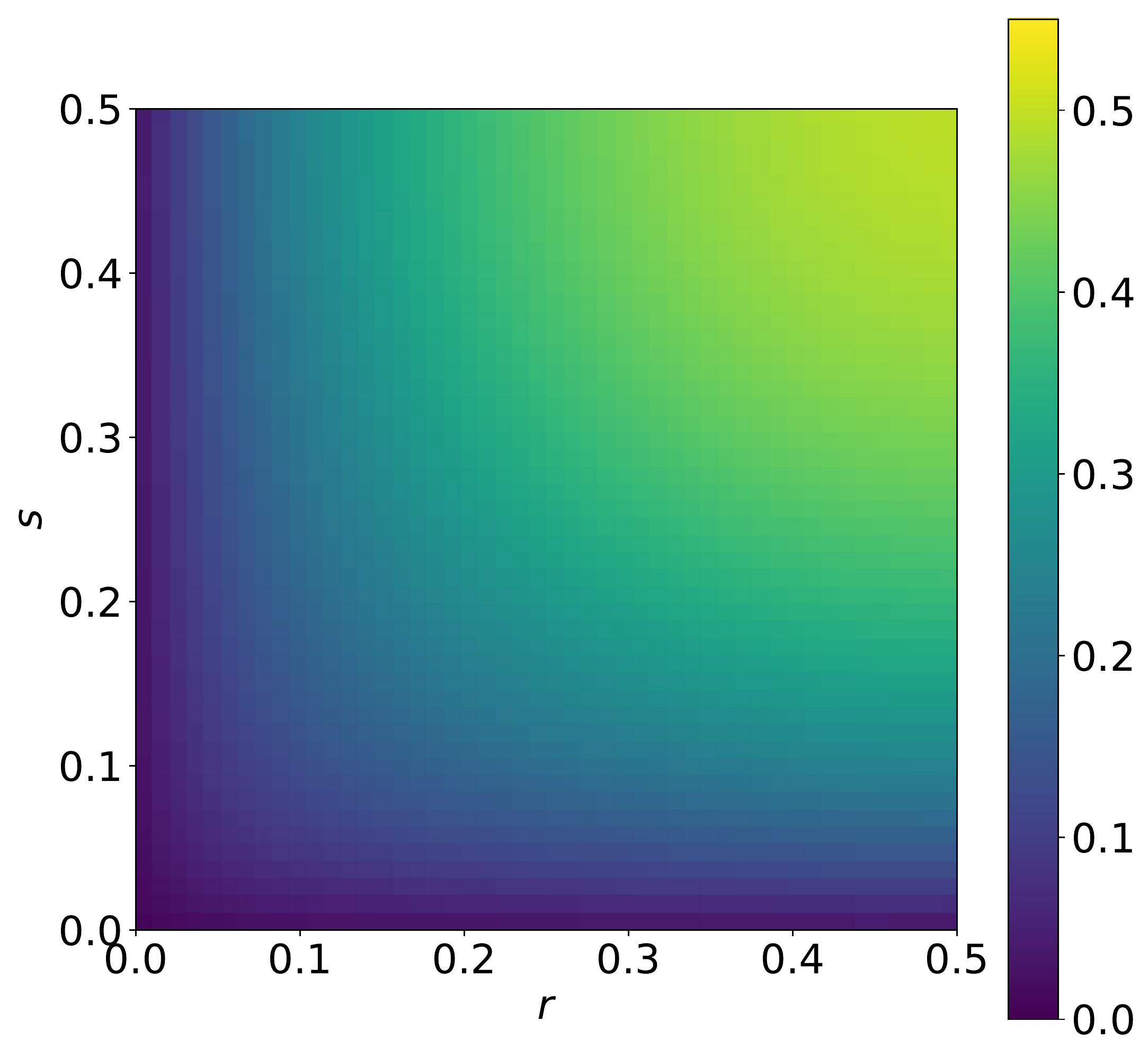}};
\draw(9.7,3.5) node{$S(r,s)-\frac{1}{n} \log\frac{A_{n+1}(rn|sn)}{A_n}$};
\draw(-0.3,3.5) node{$\frac{1}{n} \log\frac{A_{n+1}(rn|sn)}{A_n}$};
\end{tikzpicture}
\caption{Numerical verification of the factorization hypothesis for the Top-Left refinement of ASM for $n=200$. The left figure shows the exact value, the right figure shows the difference with the predicted value obtained from the factorization hypothesis.}
\label{fig_TL}
\end{center}
\end{figure}

Another interesting 2-refinement corresponds to the enumeration of ASM$_n$ with prescribed positions of the unique 1 in the first row and first column, a case also refered to as Top-Left \cite{AR13}. If $m_{1,k}=1$ and $m_{\ell,1}=1$ for $1 \le k,\ell \le \frac n2$, it exactly corresponds to the refined partition function $A_n(k|\ell)$. The factorization conjecture would then imply that the limit $\lim_{n \to \infty}\frac 1n \log A_{n+1}(rn|sn)/A_n$ is given by $S(r,s)$ in (\ref{Srs2}), with a different behaviour below or above the curve $2r+2s-rs=1$.

The exact value of $A_n(k|\ell)$ was computed in \cite{St06}, but again in a form that does not allow for an asymptotic estimate. A numerical verification is presented in Figure \ref{fig_TL}, which confirms the previous limit.

We have carried out similar checks on classes of ASM with symmetries, in particular the 1-refined partition functions $A_{\rm V}(2m+1,k)$ and $A_{\rm HT}(2m,k)$, counting respectively the vertically symmetric ASM (VSASM) of order $2m+1$ with a 1 at position $k \le m$ on the second row, and the half-turn symmetric ASM (HTSASM) of order $2m$ with a unique 1 on the first row in column $k$. Their exact expressions are both given in \cite{FS19}, from which the asymptotic behaviour can be determined analytically. In the case of VSASM, the arctic curve is the same \cite{DFL18} as for ordinary ASM and it is also expected to be the case for HTSASM. In both cases, using the ASM arctic curve we check that the factorization hypothesis is fulfilled.

\section{The tangent method revisited \label{sec5}}

This last section takes advantage of the factorization property to reformulate the tangent method in a simpler way. 
The core of the tangent method is to use a 1-refined partition function to compute a family of straight lines tangent to the arctic curve, which is in turn determined from it. In its usual form, it is done by moving the starting point of the outermost path from its original location to another point in an extension of the domain. The equation of the corresponding tangent line is then obtained by computing the likeliest entry point in the original domain, usually through the saddle point method. Here we show how the factorization hypothesis allows to compute the family of tangent lines directly from the 1-refinements, thereby avoiding the calculation of the likeliest entry point and the saddle point analysis. It has the advantage of being directly applicable to any 1-refinement including those that do not call for an extension of the original domain. 

\begin{figure}[t]
\begin{center}
\includegraphics[scale=1]{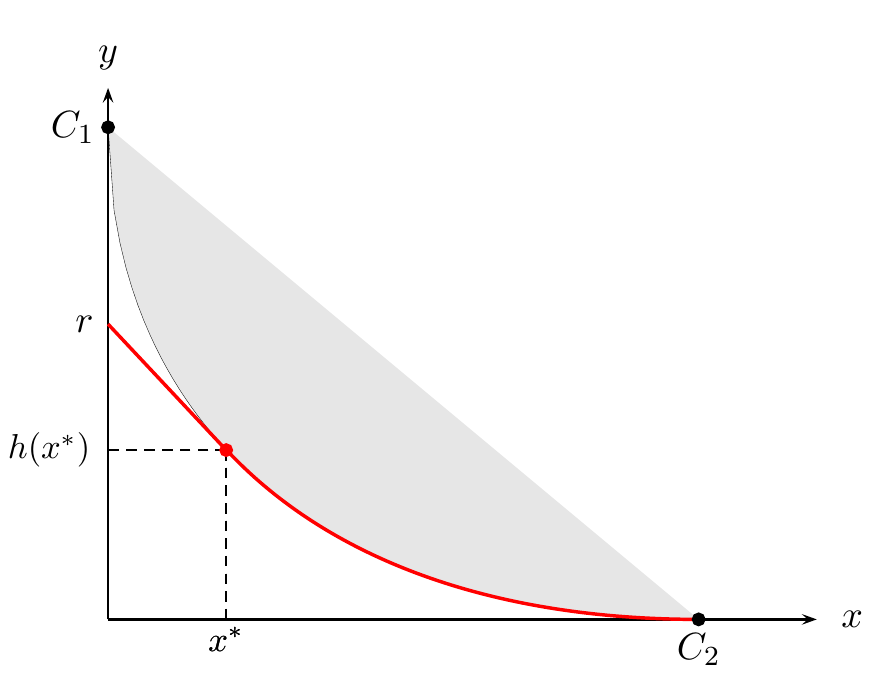}
\end{center}
\vspace{-1cm}
\caption{Typical configuration of a 1-refined calculation, shown in the scaling limit. A path starts from the point of coordinates $(0,r)$ and ends at the contact point $C_2$. The path is first a straight segment, which joins the arctic curve tangentially, at $(x^*,h(x^*))$, and follows it till $C_2$. The arctic curve is functionally given by $h(x)$.}
\label{fig8}
\end{figure}

We consider the generic situation shown in Figure \ref{fig8}, in which one intends to determine the portion of the arctic curve $y=h(x)$ between two contact points, $C_1$ and $C_2$. The white region below the arctic curve is supposed to be void of paths. The 1-refinement consists in moving the starting of the most external path from $C_1$ to the point $(0,r)$. The new external path will then look like the red curve, which comprises a straight segment hitting the arctic curve tangentially at $(x^*,h(x^*))$, with a slope $t^* = h'(x^*)$, before following the arctic curve towards $C_2$. Varying $r$ from 0 to $C_1$ yields a family of tangents to the arctic curve, which can be retrieved by solving the system of equations
\be
y = t^* x + r, \qquad \frac{\rm d}{{\rm d}r} \big[t^* x + r\big] = 0,
\label{envel}
\ee
where $t^*=t^*(r)$ is viewed as a function of $r$. The problem is to compute the function $t^*(r)$. 

%For concreteness, the method will be illustrated on the Aztec diamond with a refinement on the boundary of the original domain. The setting, drawn in the scaling limit, is pictured in Figure \ref{fig_1ref_AD}. The straight portion of the perturbed path has for equation $y = t^* (x-u) + u + 1$, where $t^*=h'(x^*)$ is seen as a function of the parameter $u$ varying in $[-1,0]$. By varying $u$, we get a family of straight lines tangent to the arctic curve, which can then be retrieved as its envelope, by solving the two following equations,
%\be
%y = t^* (x-u)+u+1, \qquad \frac{\rm d}{{\rm d}u} \big[t^* (x-u)+u+1 \big] = 0.
%\ee

Denoting by $Z_{n+1}(r)$ and $Z_n$ the 1-refined and unrefined partition functions, we consider the effective contribution in the scaling limit of the outermost red path $f^*$ as given in terms of the action associated to $f^*$,
\be
S(r) = \lim_{n\to\infty} \frac 1n \log \frac{Z_{n+1}(r)}{Z_n} = \int_0^{C_2} {\rm d}x \, L\Big(\frac{{\rm d}f^*}{{\rm d}x}\Big) = x^* L(t^*) + \int_{x^*}^{C_2} {\rm d}x \, L\big(h'(x)\big).
\label{Sr}
\ee
In this equation, the function $S(r)$ is seen as an input, obtained from independent lattice calculations, and $L$ is a function which depends on the nature and the weights of the paths in terms of which the model under consideration is formulated, as discussed in Section \ref{sec2}. 

To determine $t^*(r)$, we differentiate (\ref{Sr}) with respect to $r$,
\be
\frac{{\rm d}S(r)}{{\rm d}r} = x^* \frac{{\rm d}}{{\rm d}r} L(t^*) =  x^* L'(t^*) \, h''(x^*) \,\frac{{\rm d}x^*}{{\rm d}r}.
\ee
Since the point $(x^*,h(x^*))$ belongs to the tangent to the arctic curve at $x^*$, given by $y = t^* x + r$, it follows that
\be
r = h(x^*) - t^* x^* \quad \Longrightarrow \quad \frac{{\rm d}r}{{\rm d}x^*} = -h''(x^*) x^*.
\ee
From this, we readily obtain
\be
\frac{{\rm d}S(r)}{{\rm d}r} = - L'(t^*), \qquad {\rm or} \quad t^*(r) = (L')^{-1}\Big(\!\!-\frac{{\rm d}S(r)}{{\rm d}r}\Big),
\label{dSr}
\ee
provided the function $L'(t)$ is invertible. For the models formulated in terms of paths made of elementary steps that are weighted independently of each other, it has been proved that the function $L$ is strictly concave \cite{DGR19}. Moreover one has $-L'(t) = \log y(t)$ for a positive and strictly increasing function $y(t)$, discussed in Appendix \ref{appA}. This class of models includes the Aztec diamond and ASM, but not the general 6-vertex model which assigns each elementary step a weight that depends on the following step. Nevertheless, the explicit calculation of $L(t)$ for the general 6-vertex model shows that is also strictly concave, see Appendix \ref{appA}. So assuming the expected strict concavity of $L$, it follows that $L'(t)$ is invertible since it is then strictly decreasing on its domain. 

Let us illustrate the above procedure for ASM. The function $S(r)$, for $0 \le r \le \frac 12$, is given explicitly in (\ref{1asm}), and $-L'(t) = \log y(t) = \log\frac{1-t}{-t}$ ($t \le 0$), as recalled in Appendix \ref{appA}. The condition (\ref{dSr}) is simple and easily solved,
\be
\log\frac{1-r^2}{r(2-r)} = \log\frac{1-t^*}{-t^*} \qquad \Longrightarrow \qquad t^*(r) = \frac{r(2-r)}{2r-1}.
\ee
The two conditions (\ref{envel}) yield the parametric form of the arctic curve in a straightforward way,
\be
x = \frac{(2r-1)^2}{2(1-r+r^2)}, \quad y = \frac{3r^2}{2(1-r+r^2)}, \qquad 0 \le r \le \frac 12.
\ee
Eliminating $r$, we recover the ellipse of equation $x(1-x) + y(1-y) + xy = \frac 14$, found in \cite{CP10a}. 

\section{Conclusion}

We have proposed a factorization property for models which have a formulation in terms of paths, possibly interacting. The property states that, in the scaling limit, the contribution of the few most external paths to the partition function factorizes from that of the bulk paths, responsible for the formation of the arctic curve. Moreover the contribution of each external path is given by the action associated with a certain Lagrangean function $L$, and can be made fully explicit when the arctic curve is known. The factorization has been verified in full generality for tilings of Aztec diamonds, and for certain refinements of alternating sign matrices. Moreover, many numerical checks\footnote{The only numerical part in these verifications is the evaluation of fairly complicated integrals.}, not reported here, based on the 1-refinements considered in \cite{CPZ10,CP10b}, have been successfully carried out for the 6-vertex model, in the disordered and antiferroelectric phases (and for values of $\Delta$ up to $0.998$). 

That the factorization holds in cases where the paths interact may sound surprising, since the function $L$ is defined for an isolated path, and is therefore blind to all forms of interactions present in the model. On the other hand, at microscopic scales, the external paths are expected to stay well away from each other, and are unaffected by the interactions (unless these are long ranged, which is usually not the case).

%An interesting question remains open for ASM with symmetry constraints when the symmetries are not shared by the arctic curve of unconstrained ASM, {\red which could happen if we consider weights which are themselves not symmetric}. Indeed the function $L$ cannot incorporate the symmetries as it is defined for isolated paths. In such a case, the factorization could presumably fail.

Exact expressions for multirefinements can be extremely hard to obtain on the lattice. Even when an explicit formula is known, evaluating its asymptotics may be challenging. In this regard, the factorization hypothesis provides a remarkably simple, efficient and explicit way to obtain the asymptotics of multirefined partition functions. As yet it remains however conjectural but verified in many instances.

%%%%%%%%%%%%%%%%%%%%%%%%%%%%%%%%%%%%%%%%%%%%%%%%%%%%%%%%%%%%%%%%%%%%%%%%%%%%%%%%%%%%%%%%%%%%%%%%%%%%%%%%%

\appendix
\section{Asymptotics for directed paths}
\label{appA}

This appendix collects some useful results about the asymptotic enumeration of weighted directed paths, and in particular the functions $L$ which appear throughout the text. Within the context of the tangent method, these functions $L$ were introduced in \cite{DGR19}. We restrict here to paths in $\Z^2$.

An important and useful class of paths is when the paths are made of elementary steps $\vec s_i$ taken from a finite set $S = \{\vec s_i = (u_i,v_i)\}$, each step being assigned a weight $w_i > 0$. The weight of a path is equal to the product of the weights of its elementary steps. A basic question is to compute the weighted sum $Z_{p,q}$ of all paths from $(0,0)$ to $(p,q)$. The generating function of these numbers is given by
\be
G(x,y) = \sum _{p,q} Z_{p,q}\, x^p y^q = \frac 1{1 - P(x,y)}, \qquad P(x,y) = \sum_i w_i \, x^{u_i} y^{v_i}.
\ee
For directed paths, all numbers $Z_{p,q}$ are finite.

We are especially interested in the asymptotic value of $Z_{p,q}$ when $p,q$ are proportional to a large $n$, say $p=rn,\,q=sn$. Then the following limit exists,
\be
L(t) = \lim_{n \to \infty} \frac 1{rn} \log Z_{rn,sn}, \qquad t = \frac sr.
\label{Llim}
\ee
Moreover it can be very explicitly written as \cite{PW08}\
\be
L(t) = -\log x(t) - t \log y(t),
\ee
where $x(t)$ and $y(t)$ are the unique positive and real solutions of the following system,
\be
P(x,y) = 1, \qquad t x \, \partial_xP(x,y) = y \, \partial_yP(x,y).
\label{syst}
\ee

For the two examples worked out in the text, namely Aztec diamonds and ASM, these functions are given as follows,
{\small
\bea
&&
\hspace{-7mm}
\psset{unit=5mm}
\begin{tikzpicture}[scale=0.5,baseline=4mm]
\draw[gray] (0,0)--(2,0);
\draw[gray] (0,1)--(2,1);
\draw[gray] (0,2)--(2,2);
\draw[gray] (0,0)--(0,2);
\draw[gray] (1,0)--(1,2);
\draw[gray] (2,0)--(2,2);
\draw[->,>=latex,thick,red] (0,1) -- (1,2);
\draw[->,>=latex,thick,red] (0,1)--(1,0);
\draw[->,>=latex,red,thick] (0,1)--(2,1) node[xshift=-0.2cm,below,black]{$w$};
\end{tikzpicture}
\qquad x(t) = \frac{\sqrt{1+w} - \sqrt{1+wt^2}}{w\sqrt{1-t^2}}, \quad y(t) = \frac{t\sqrt{1+w} + \sqrt{1+wt^2}}{\sqrt{1-t^2}}, \quad -1 < t < 1.
\label{eq_xy2} \\
\noalign{\medskip}
&& 
\hspace{-7mm}
\psset{unit=5mm}
\begin{tikzpicture}[scale=0.5,baseline=4mm]
\draw[gray] (0,0)--(2,0);
\draw[gray] (0,1)--(2,1);
\draw[gray] (0,2)--(2,2);
\draw[gray] (0,0)--(0,2);
\draw[gray] (1,0)--(1,2);
\draw[gray] (2,0)--(2,2);
\draw[->,>=latex,thick,red] (0,1) -- (1,1);
\draw[->,>=latex,thick,red] (0,1)--(0,0);
\end{tikzpicture}
\qquad x(t) = \frac 1{1-t}, \quad y(t) = -\frac {1-t}t, \qquad t < 0,
\label{eq_xy1}
\eea
}
The elementary steps are indicated by the red arrows; all steps have a weight 1 except the step $(2,0)$ in the first example, which has weight $w$.

As a consequence of $x,y$ solving the system (\ref{syst}), it has been shown in \cite{DGR19} that the three functions $x(t),y(t),L(t)$ are $C^\infty$ on their domain; moreover $y(t)$ is strictly increasing. On account of $L'(t) = -\log y(t)$, it follows that $L''(t) = -\frac {y'(t)}{y(t)} < 0$, and so $L(t)$ is strictly concave.

In the general 6-vertex model, the isolated paths are made of the same two steps $\vec s_1 = (1,0)$ and $\vec s_2 = (0,-1)$ as in the second example above, but they are weighted in a different way. Indeed the weight of a step depends on what the next step is, because the 6-vertex model primarily attaches weights to vertices, seen as points in between two successive steps. Two successive and identical steps (a straight) gets a weight $w_1$ and two successive but different steps (a turn) gets a weight $w_2$ ($w_1$ and $w_2$ are respectively equal to $\frac ba$ and $\frac ca$ in the usual notations of the 6-vertex model). The reference \cite{CS16} gives the general expression for the weighted sum of all paths from $(0,0)$ to $(p,q)$, $q \le 0 \le p$,
\be
Z_{p,q}(w_1,w_2) = w_1^{p-q+1} \sum_{\ell=0}^{\min(p,-q)} {p \choose \ell}{-q \choose \ell} \tau^{2\ell+1}, \qquad \tau =\frac {w_2}{w_1}.
\ee
In the scaling limit, $p=rn,q=sn$, the limit (\ref{Llim}) gives
\be
L(t;w_1,w_2) = (1-t) \log w_1 + (1-t) L\Big(\frac{1+t}{1-t};w=\tau^2-1\Big), \qquad t = \frac sr \le 0,
\ee
where in the r.h.s, $L(t;w)$ is the function pertaining to the first example (\ref{eq_xy2}) above. In the case $w_1=w_2=1$, the above general expression reduces to the function $L(t;1,1) = (1 - t) \log(1 - t) + t \log |t|$ of the second example (\ref{eq_xy1}), as used in Section \ref{sec4}. On account of the identity
\be
\frac{{\rm d}^2}{{\rm d}t^2}\Big[(1-t)L\Big(\frac{1+t}{1-t}\Big)\Big] = \frac{4}{(1-t)^3}L''\Big(\frac{1+t}{1-t}\Big),
\ee
valid for any function $L$, the strict concavity of $L(t;w)$ implies that of $L(t;w_1,w_2)$. 

%The $x$-enumeration of ASM is recovered for $w_1=1$ and $w_2=\sqrt{x}$. 

%%%%%%%%%%%%%%%%%%%%%%%%%%%%%%%%%%%%%%%%%%%%%%%%%%%%%%%%%%%%%%%%%%%%%%%%%%%%%%%%%%%%%%%%%%%%%%%%%%%%%%%%%
\section{Technicalities}
\label{appB}

This short Appendix is devoted to the proof of the property, used in Section \ref{multi}, that for the function $S(r,s)$ in (\ref{Srs}), the difference $S(r,s)-S(r,s')$ is increasing in $r$ for any fixed $s > s'$, and strictly increasing if $rs > \frac 1{1+w}$. As the function $S(r,s)$ has a different functional form according to the value of $rs$, we distiguish three separate cases. The various relations between the function $x(t),y(t)$ and $L(t)$ have been recalled in Appendix A.

For $rs' < rs \le \frac 1{1+w}$, $S(r,s)$ is equal to $\sigma(r) + \sigma(s)$, with
\be
\sigma(x) = (x+1) \log{(x+1)} - x \log x +\frac 12 \log{(1+w)}, \qquad \sigma'(x) = \log\frac{x+1}x.
\ee
It follows that $S(r,s) - S(r,s') = \sigma(s) - \sigma(s')$ is constant in $r$.

For $rs' \le \frac 1{1+w} < rs$, we have 
\be
S(r,s) - S(r,s') = (r+s+2) L(t) - \sigma(r) - \sigma(s'), \qquad t = \frac{s-r}{r+s+2}.
\ee
By using that $L(t) = -\log x(t) - t \log y(t)$ and the fact that $L'(t) = -\log y(t)$, we obtain that the derivative with respect to $r$ simplifies to
\be
\frac {\rm d}{{\rm d}r} \big[S(r,s) - S(r,s')\big] = \log \frac{r \, y(t)}{(r+1)\,x(t)}.
\ee
The explicit forms of the functions $x(t)$ and $y(t)$ show that the argument of the logarithm is strictly larger than 1 for $r > \frac 1{s(1+w)}$ (it is exactly 1 for $r = \frac 1{s(1+w)}$ and tends to 1, from above, when $r \to +\infty$). Therefore the difference $S(r,s) - S(r,s')$ is strictly increasing. 

In the last case, namely for $\frac 1{1+w} < rs' < rs$, we have
\be
S(r,s) - S(r,s') = (r+s+2) L(t) - (r+s'+2) L(t'), \qquad t' = \frac{s'-r}{r+s'+2}.
\ee
The calculation of the previous paragraph readily yields
\be
\frac {\rm d}{{\rm d}r} \big[S(r,s) - S(r,s')\big] = \log \frac{y(t) \, x(t')}{x(t) \, y(t')} > 0.
\ee
The inequality follows from the ratio $y(t)/x(t)$ being itself a strictly increasing function on its open domain $]-1,1[$,
\be
\frac {\rm d}{{\rm d}t} \frac{y(t)}{x(t)} = \frac{y'(t) x(t) - y(t) x'(t)}{x^2(t)} = (1+t) \frac{y'(t)}{x(t)} > 0.
\ee
The fact that $s'<s$ implies the strict inequality $t'<t$ between the two slopes concludes the argument and shows that $S(r,s) - S(r,s')$ is strictly increasing in $r$.

\section*{Acknowledgments}
This work was supported by the Fonds de la Recherche Scientifique\,-\,FNRS and the Fonds Wetenschappelijk 
Onderzoek\,-Vlaanderen (FWO) under EOS project no 30889451. P.R. is a Senior Research Associate of FRS-FNRS (Belgian Fund for Scientific Research).

%%%%%%%%%%%%%%%%%%%%%%%%%%%%%%%%%%%%%%%%%%%%%%%%%%%%%%%%%%%%%%%%%%%%%%%%%%

\end{document}